\documentclass[review]{elsarticle}

\usepackage{lineno,hyperref,multirow}
\modulolinenumbers[5]

\usepackage[left=2.4cm,top=2.2cm,right=2.3cm,bottom=2.4cm,nohead]{geometry}
\usepackage{subcaption}
 \biboptions{sort&compress}

\usepackage{graphicx,dsfont,bm,bbold,float,xcolor,hyperref,amsmath}
\usepackage{subcaption}

\newcommand{\tr}{\text{\,tr\,}}
\newcommand{\pf}{\text{\,Pf\,}}

\newcommand{\diag}{\text{diag\,}}

\newcommand{\1}{\mathds{1}_n}
\newcommand{\bSg}{\boldsymbol{\Sigma}}

\newcommand{\LL}{\mathbb{l}}
\newcommand{\UL}{\mathbb{u}}
\newcommand{\Bt}{\text{B}}
\newcommand{\bG}{{\bf G}}
\newcommand{\bH}{{\bf H}}
\newcommand{\bA}{{\bf A}}
\newcommand{\bB}{{\bf B}}
\newcommand{\bK}{{\bf K}}

\newcommand{\bU}{{\bf U}}
\newcommand{\bW}{{\bf W}}
\newcommand{\bY}{{\bf Y}}
\newcommand{\erf}{\text{erf}}
\newcommand{\bbL}{\mathbb{L}}

\begin{document}

\begin{frontmatter}

\title{Gap probabilities and densities of extreme eigenvalues of random matrices: Exact results}

\author{Santosh Kumar}
\address{Department of Physics, Shiv Nadar University, Gautam Buddha Nagar, Uttar Pradesh - 201314, India}
\ead{skumar.physics@gmail.com}




\begin{abstract}
We derive exact results for gap probabilities, as well as densities of extreme eigenvalues for six complex random matrix ensembles of fundamental importance. These are Gauss-Wigner, Laguerre-Wishart, Cauchy-Lorentz (two variants), Jacobi-MANOVA and Bures-Hall (trace unrestricted). We deal with both correlated and uncorrelated cases. Extensive Monte-Carlo simulations based on explicit matrix models or Dyson's log-gas formalism have also been performed, which corroborate all analytical results.
\end{abstract}


\end{frontmatter}



\section{Introduction}

Given a random matrix ensemble, the question regarding probabilities of gaps in its eigenvalue spectrum is of fundamental importance. The investigation of gap probabilities is not only interesting from a general mathematical point of view, but also because of its connection with several interesting problems and concrete applications. Naturally there has been a plethora of studies which deal directly or indirectly with this question~\cite{AE2006,BD2007,FN2012,Edelman1989,WF2000,WFC2000,Krasovsky2004,APS2009,LL2013,JDM2008,TW1994,TW1996,Edelman1991,May1972,KL2005,DM2008,LTBM2008,Forrester2007,CTKV2002,Burel2002,ZNYZY2008,NZYY2008,MBL2008,CLZ2010,AV2011,Nadler2011,WTDM2012, KSRZ2012,Johnstone2001,Johnstone2008,RKC2012,KD2008,ZCW2009,WG2013,WG2014,WKG2015,AGKWW,WAGKW}. For example, in the case of a Hessian matrix drawn from a Gaussian ensemble, which can model a random multi-field potential or a landscape, the gap probability serves as an estimator for the fraction of local minima amongst the stationary points of the landscape~\cite{AE2006,BD2007,FN2012}. As application to geometry and random topology, gap probability has been used to calculate the intrinsic volume of the set of Frobenius norm one singular random matrices, and computation of certain class of Betti numbers~\cite{LL2013}. In~\cite{Krasovsky2004} gap probability has been shown to be related to the asymptotics of orthogonal polynomials defined on an arc of unit circle. In~\cite{JDM2008} a connection has been established with the expected number of minima of a random polynomial. Gap probabilities are also relevant in deciding physical stability in dynamical systems and ecosystems~\cite{May1972,KL2005}. 

Intimately related to gap probabilities is the topic of extreme eigenvalues, which has been of considerable interest to researchers in several fields~\cite{FN2012,TW1994,TW1996,Edelman1991,DM2008,FPNFD2012,LTBM2008,Forrester2007,CTKV2002,Burel2002,ZNYZY2008,NZYY2008,MBL2008,CLZ2010,AV2011,Nadler2011,WTDM2012, KSRZ2012,Johnstone2001,Johnstone2008,RKC2012,KD2008,ZCW2009,WG2013,WG2014,WKG2015,AGKWW,WAGKW}. One of the notable examples is the celebrated Tracy-Widom distribution for extreme eigenvalues~\cite{TW1994,TW1996}, which emerges in seemingly unrelated problems, such as mesoscopic fluctuations of excitation gap in quantum dots~\cite{VBAB2001,OSF2001}, distribution of the pseudo-crtical temperature in mean-field spin glasses~\cite{CDZ2011}, height fluctuations of non-intersecting Brownian motions~\cite{FMS2011,Liechty2012}, growth models~\cite{PS2000,Johansson2000,GTW2001,IS2004}, sequence alignment~\cite{MN2004}, random permutations~\cite{BDJ1999}, finance~\cite{BBP2007}, etc. In the estimation of performance of multiple channel communication systems the statistics of extreme singular values (square root of eigenvalues) of channel matrix plays a crucial role~\cite{CTKV2002,Burel2002,ZNYZY2008,NZYY2008,ZCW2009}. In the problem of quantum entanglement the extreme Schmidt eigenvalues carry important information about the degree of entanglement~\cite{MBL2008,CLZ2010,AV2011}. In finance, the optimal portfolio is related to the extreme eigenvalues of the correlation matrix~\cite{Markowitz1959}. Extreme eigenvalues also lead to determination of metric such as condition number~\cite{Edelman1989,RVA2004,MMSN2010}, which is a measure of the relative conditioning (or rank-deficiency) of a matrix and has numerous applications in a variety of contexts, for example multiple channel communication systems, linear detection and classical linear algebra. Moreover, in recent years the large deviations of extreme eigenvalues of random matrices have been studied in several contexts~\cite{MV2009,TM2013,MSVV2012,DM2006,VMB2007} which range from principal component analysis to Wigner time-delay distribution in chaotic cavities. A recent survey can be found in~\cite{MS2014}. Large deviation studies relate to yet another riveting topic of extreme or rare events~\cite{Gumbel2004,AJK2006}.

The case of matrices with large dimensions has been exhaustively studied for extreme eigenvalues, while for finite dimensionality case there have been relatively fewer studies with analysis remaining concentrated mostly around Laguerre-Wishart ensembles. Our aim here is to derive finite dimensionality exact results for gap probabilities, and probability density functions for extreme eigenvalues. We cover the following ensembles of complex matrices which are of central importance to random matrix theory and have wide range of applications:
\begin{enumerate}
\item Gauss-Wigner Ensemble
\item Laguerre-Wishart Ensemble
\item Cauchy-Lorentz Ensemble (Variant I)
\item Cauchy-Lorentz Ensemble (Variant II)
\item Jacobi-MANOVA Ensemble
\item Bures-Hall Ensemble (trace unrestricted)
\end{enumerate}
We consider both correlated, and uncorrelated cases of these ensembles. In the case of uncorrelated ensembles the underlying matrix probability measure is rotationally invariant. On the other hand, associated with correlated ensembles are rotationally noninvariant probability measures which are comparatively difficult to handle. As we find below, these ensembles fall broadly under two categories: the ones with joint density of eigenvalues involving product of two determinants, and the others with joint density involving product of a determinant and a Pfaffian~~\cite{Mehta2004}. Therefore, we also provide general results based on generalizations of Andr\'{e}ief's formula~\cite{Andreief1883,KG2010a,Kumar2015} and de Brujin's formula~\cite{Brujin1955,Kieburg2012}. 

The remaining of the paper is organized as follows. We focus on the generic setup in Sec~\ref{SecGen} and derive general results for the two categories of ensembles indicated above. Sections~\ref{SecWig} to \ref{SecBH} deal with explicit results for the six (times two) random matrix ensembles listed above. We conclude in section~\ref{SecConc} with a brief summary and outlook. 


\section{General results}
\label{SecGen}

\subsection{Preliminaries}
\label{Pre}

Consider the joint probability density function (JPDF) $P(\{\lambda\})$ of $n$ eigenvalues $\lambda_1,...,\lambda_n$ associated with certain random matrix ensemble, such that
\begin{equation}
\int_\LL^\UL d\lambda_1\cdots\int_\LL^\UL d\lambda_n\, P(\{\lambda\})=1.
\end{equation}
Here $\LL$ and $\UL$ are the lower and upper integration limits and set the domain for the eigenvalues. In the cases when $P(\{\lambda\})$ has an associated easy to implement explicit matrix model, one can generate the eigenvalues by diagonalization. While for any well behaved JPDF, with or without any explicit matrix model, one can always generate the eigenvalues using the Dyson's log gas formalism~\cite{GPPS2003,Kumar2014}. Given this, one is naturally interested in some metric to determine the fraction of eigenvalues which lie in a certain interval within the domain of the eigenvalues. This is equivalent to figuring out the probability that a certain interval is devoid of any eigenvalues, and hence one is led to the study of gap probability. A closely related problem is the investigation of behavior of the smallest or the largest eigenvalue, for which the full information is provided by the corresponding probability densities. In this section we formally define these quantities and afterwards work out the expressions for two kinds of structures for the matrix ensembles which are very generic and cover a number of important cases.

We will calculate the gap probability
\begin{equation}
\label{Ers}
E(r,s)=\int\limits_{(\LL,r)\cup(s,\UL)} d\lambda_1\cdots\int\limits_{(\LL,r)\cup(s,\UL)} d\lambda_n\, P(\{\lambda\}),
\end{equation}
which refers to the probability of finding no eigenvalue in $(r,s)$, such that $\LL\leq r\leq s\leq \UL$. We also examine the probability that all the eigenvalues lie in $[r,s]$:
\begin{equation}
\label{tErs}
\widetilde{E}(r,s)=\int_r^s d\lambda_1\cdots\int_r^s d\lambda_n\, P(\{\lambda\}).
\end{equation}
We should underline that 
\begin{equation}
\widetilde{E}(r,s)\neq 1-E(r,s),
\end{equation}
except when $n=1$. We observe that $\widetilde{E}(r,s)$ is same as the gap probability that there are no eigenvalues in either of the intervals $(\LL,r)$ and $(s,\UL)$. We will refer to it as the double gap probability.

Of particular interest are the gap probabilities $E(\LL,x)$ and $E(x,\UL)$, with $\LL\le x \le \UL$. Clearly 
\begin{equation}
E(\LL,x)=\widetilde{E}(x,\UL)= \int_x^\UL d\lambda_1\cdots\int_x^\UL  d\lambda_n\, P(\{\lambda\})
\end{equation}
gives the probability that there are no eigenvalues between $\LL$ and $x$, which is same as the probability that all eigenvalues are greater than or equal to $x$. As such, it refers to the survival function (SF) or reliability function of the smallest eigenvalue. On the other hand
\begin{equation}
E(x,\UL)=\widetilde{E}(\LL,x)= \int_\LL^x d\lambda_1\cdots\int_\LL^x  d\lambda_n\, P(\{\lambda\})
\end{equation}
gives the probability that there are no eigenvalue between $x$ and $\UL$ or that all eigenvalues are less than or equal to $x$. Therefore, it is the cumulative distribution function (CDF) of the largest eigenvalue. Survival function and cumulative distribution function are related as SF$=1-$CDF.

The probability density function (PDF) of the smallest eigenvalue, and that of the largest eigenvalue are obtained respectively as
\begin{equation}
\label{pS}
p_{\mathbb{S}}(x)=-\frac{\partial E(\LL,x)}{\partial x}=-\frac{\partial \widetilde{E}(x,\UL)}{\partial x},
\end{equation}
and
\begin{equation}
\label{pL}
p_{\mathbb{L}}(x)=\frac{\partial E(x,\UL)}{\partial x}=\frac{\partial \widetilde{E}(\LL,x)}{\partial x}.
\end{equation}
We may also evaluate the joint probability density of the smallest and the largest eigenvalues using $\widetilde{E}(r,s)$ as
\begin{equation}
\label{pSL}
p_{\mathbb{SL}}(r,s)=-\frac{\partial^2 \widetilde{E}(r,s)}{\partial r\partial s}.
\end{equation}
With the information of $p_{\mathbb{SL}}(r,s)$ we can examine the statistics of quantities which involve the smallest eigenvalue as well as the largest eigenvalue. For instance, we can compute the density of condition number.

As already indicated, in the matrix ensembles that we pursue in the following sections, the JPDF exhibits two kinds of structure. The first one has a biorthogonal form of Borodin type~\cite{Borodin1998} and involves product of two determinants. The second one involves, on the other hand, the product of a determinant and a Pfaffian. We will refer to these as ensembles of Type I and Type II, respectively. We present the general results for these ensembles below, and explicit results for the particular cases are given in the subsequent sections.

\subsection{Type I Ensembles}
\label{SecType1}

We consider in this category, the ensembles which involve product of two determinants, namely
\begin{equation}
\label{Type1}
P(\{\lambda\})=C\prod_{l=1}^nw(\lambda_l)\cdot |f_j(\lambda_k)|_{j,k=1,...,n}\,|g_j(\lambda_k)|_{j,k=1,...,n}.
\end{equation}
Here we use the notation $|\cdot|$ to represent determinant. We note that the product involving the weight function $w(\lambda)$ can be absorbed within the determinants. This is clearly of Borodin's biorthogonal form~\cite{Borodin1998}. We will deal with correlated ensembles for which we find $f_j(\lambda_k)=\lambda_k^{j-1}$ which makes the corresponding determinant the Vandermonde determinant:
\begin{equation}
\Delta_n(\{\lambda\})=|\lambda_k^{j-1}|_{j,k=1,...,n}=\prod_{j>k}(\lambda_j-\lambda_k).
\end{equation}
Eq.~\eqref{Type1} also includes the case of classical unitary ensembles which arise when in addition to $f_j(\lambda_k)$, $g_j(\lambda_k)=\lambda_k^{j-1}$ as well. This situation is encountered in the case of uncorrelated ensembles.

For this class of ensembles Andr\'{e}ief's integration formula~\cite{Andreief1883,KG2010a,Kumar2015} yields at once the expression for the normalization factor, as well as the gap probabilities $E(r,s)$ and $\widetilde{E}(r,s)$. The inverse of normalization factor (partition function) is found to be
\begin{equation}
C^{-1}=n!\,|h_{j,k}|_{j,k=1,...,n}
\end{equation}
where
\begin{equation}
h_{j,k}=\int_\LL^\UL d\lambda\, w(\lambda) f_j(\lambda) g_k(\lambda)
\end{equation}

The gap probability is obtained as
\begin{equation}
\label{Ers1}
E(r,s)=n! \,C |\chi_{j,k}(r,s)|_{j,k=1,...,n},
\end{equation}
with the kernel $\chi_{j,k}(r,s)$ given by
\begin{equation}
\chi_{j,k}(r,s)=\int\limits_{(\LL,r)\cup(s,\UL)}d\lambda\, w(\lambda) f_j(\lambda) g_k(\lambda)
\end{equation}
The expressions for $E(\LL,s)$ and $E(r,\UL)$ can be read easily from Eq.~\eqref{Ers1} with the kernel $\chi_{j,k}$ getting simplified respectively to
\begin{equation}
\chi_{j,k}(\LL,s)=\int_s^\UL d\lambda\, w(\lambda) f_j(\lambda) g_k(\lambda),
\end{equation}
and
\begin{equation}
\chi_{j,k}(r,\UL)=\int_\LL^r d\lambda\, w(\lambda) f_j(\lambda) g_k(\lambda).
\end{equation}
Similarly, $\widetilde{E}(r,s)$ has the expression
\begin{equation}
\label{tErs1}
\widetilde{E}(r,s)=n! \,C |\widetilde{\chi}_{j,k}(r,s)|_{j,k=1,...,n},
\end{equation}
where the kernel $\widetilde{\chi}_{j,k}(r,s)$ is given by
\begin{equation}
\widetilde{\chi}_{j,k}(r,s)=\int_r^s d\lambda\, w(\lambda) f_j(\lambda) g_k(\lambda).
\end{equation}

The expression for density of the smallest eigenvalue is obtained using Eq.~\eqref{pS} upon differentiating the determinantal expression as
 \begin{align}
\label{pS1}
p_{\mathbb{S}}(x)&=n!\,C\,\sum_{i=1}^n |\phi_{j,k}^{(i)}(x)|_{j,k=1,...,n},
\end{align}
where
\begin{equation}
\phi_{j,k}^{(i)}(x)
=\begin{cases}
w(x) f_j(x) g_k(x), & j=i,\\
 \chi_{j,k}(\LL,x), & j\neq i.
 \end{cases}
 \end{equation}
In a similar manner Eq.~\eqref{pL} gives the density of the largest eigenvalue as
\begin{align}
\label{pL1}
p_{\mathbb{L}}(x)&=n!\,C\,\sum_{i=1}^n |\psi_{j,k}^{(i)}(x)|_{j,k=1,...,n},
\end{align}
with
\begin{equation}
\psi_{j,k}^{(i)}(x)
=\begin{cases}
w(x) f_j(x) g_k(x), & j=i,\\
 \chi_{j,k}(x,\UL), & j\neq i.
 \end{cases}
 \end{equation}
It is to be noted here that $\chi_{j,k}(\LL,x)=\widetilde{\chi}_{j,k}(x,\UL)$, and $\chi_{j,k}(x,\UL)=\widetilde{\chi}_{j,k}(\LL,x)$.


\subsection{Type II Ensembles}
\label{SecType2}

We consider in this category the ensembles whose joint probability density involve product of a Pfaffian (\pf) and a determinant, i.e.
\begin{equation}
\label{Type2}
P(\{\lambda\})=C\prod_{l=1}^nw(\lambda_l)\cdot \pf[f(\lambda_j,\lambda_k)]_{j,k=1,...,n} |g_j(\lambda_k)|_{j,k=1,...,n} \,.
\end{equation}
The kernel $f(\lambda,\mu)$ is antisymmetric in $\lambda$ and $\mu$, i.e., $f(\lambda,\mu)=-f(\mu,\lambda)$. Ensembles of above type have been encountered in the study of Wilson random matrix theory~\cite{Kieburg2012}. Moreover, the JPDF of eigenvalues for unrestricted trace Bures-Hall ensemble can also be cast in the above form~\cite{FK2015}; see section~\ref{SecBH} ahead. The normalization factor, as well as the gap probability expressions can be obtained using a generalization of de Bruijn integration theorem~\cite{Brujin1955,Kieburg2012}. 

Let us define for convenience
\begin{equation}
\label{DefN}
N=\begin{cases}
~~n, & n \text{ even},\\
n+1, & n \text{ odd}.
\end{cases}
\end{equation}
We then have the expression for the partition function as
\begin{align}
\label{Ct2}
C^{-1}= n! \pf[h_{j,k}]_{j,k=1,...,N},
\end{align}
where
\begin{align}
\label{hjke}
\nonumber
h_{j,k}&=\int_\LL^\UL d\lambda \int_\LL^\UL d\mu\, w(\lambda)w(\mu) f(\lambda,\mu) g_j(\lambda) g_k(\mu)\\
&=\frac{1}{2}\int_\LL^\UL d\lambda \int_\LL^\UL d\mu\, w(\lambda)w(\mu)f(\lambda,\mu) \, [g_j(\lambda) g_k(\mu) -g_k(\lambda)g_j(\mu)]
\end{align}
when $n$ is even, and in addition
\begin{equation}
\label{hjko}
h_{j,n+1}=-h_{n+1,j}= (1-\delta_{j,n+1})\int_\LL^\UL d\lambda \, w(\lambda) g(\lambda)
\end{equation}
when $n$ is odd. Here, $\delta_{i,j}$ is the Kronecker delta and therefore the factor $(1-\delta_{j,n+1}) $ ensures that $h_{n+1,n+1}=0$.
We found that the second expression in Eq.~\eqref{hjke} is more stable if used for numerical evaluation.

The gap probability can also be written in terms of a Pfaffian as
\begin{equation}
\label{Ers2}
E(r,s)=n!\, C\pf[\chi_{j,k}(r,s)]_{j,k=1,...,N},
\end{equation}
with the kernel $\chi_{j,k}(r,s)$ given by
\begin{align}
\nonumber
\chi_{j,k}(r,s)&=\int\limits_{(\LL,r)\cup(s,\UL)} d\lambda \int\limits_{(\LL,r)\cup(s,\UL)} d\mu\, w(\lambda)w(\mu) f(\lambda,\mu) g_j(\lambda) g_k(\mu) \\
&=\frac{1}{2}\int\limits_{(\LL,r)\cup(s,\UL)} d\lambda \int\limits_{(\LL,r)\cup(s,\UL)} d\mu \,w(\lambda)w(\mu)f(\lambda,\mu) \, [g_j(\lambda) g_k(\mu) -g_k(\lambda)g_j(\mu)]
\end{align}
when $n$ is even. When $n$ is odd, additionally we have
\begin{equation}
\chi_{j,n+1}(r,s)=-\chi_{n+1,j}(r,s)=(1-\delta_{j,n+1}) \int\limits_{(\LL,r)\cup(s,\UL)} d\lambda \, w(\lambda) g_j(\lambda).
\end{equation}
Similarly, the double gap probability also turns out to be a Pfaffian:
\begin{equation}
\label{tErs2}
\widetilde{E}(r,s)=n! \,C\pf[\widetilde{\chi}_{j,k}(r,s)]_{j,k=1,...,N}.
\end{equation}
Here the kernel $\widetilde{\chi}_{j,k}(r,s)$ is
\begin{align}
\widetilde{\chi}_{j,k}(r,s)&=\int_r^s d\lambda \int_r^s d\mu\, w(\lambda)w(\mu) f(\lambda,\mu) g_j(\lambda) g_k(\mu)\\
&=\frac{1}{2}\int_r^s d\lambda \int_r^s d\mu\,w(\lambda)w(\mu) f(\lambda,\mu) \, [g_j(\lambda) g_k(\mu) -g_k(\lambda)g_j(\mu)]
\end{align}
when $n$ is even, and in addition
\begin{equation}
\widetilde{\chi}_{j,n+1}(r,s)=-\widetilde{\chi}_{n+1,j}(r,s)=(1-\delta_{j,n+1}) \int_r^s d\lambda \, w(\lambda) g_j(\lambda)
\end{equation}
when $n$ is odd. 

The expression for density of the smallest eigenvalue is obtained using Eq.~\eqref{pS} by differentiating the Pfaffian-expression as
 \begin{align}
\label{pS2}
p_{\mathbb{S}}(x)&=\frac{1}{2}n!\,C\,\pf^{-1}[\chi_{j,k}(\LL,x)]_{j,k=1,...,N}\sum_{i=1}^n |\phi_{j,k}^{(i)}(x)|_{j,k=1,...,N},
\end{align}
with
\begin{equation}
\phi_{j,k}^{(i)}(x)
=\begin{cases}
-\dfrac{\partial}{\partial x} \chi_{j,k}(\LL,x), & j=i,\\
 \chi_{j,k}(\LL,x), & j\neq i.
 \end{cases}
 \end{equation}
Likewise, the expression for density of the largest eigenvalue is given by
 \begin{align}
\label{pL2}
p_\bbL(x)&=\frac{1}{2}n!\,C\,\pf^{-1}[\chi_{j,k}(x,\UL)]_{j,k=1,...,N}\sum_{i=1}^n |\psi_{j,k}^{(i)}(x)|_{j,k=1,...,N},
\end{align}
where
\begin{equation}
\psi_{j,k}^{(i)}(x)
=\begin{cases}
\dfrac{\partial}{\partial x} \chi_{j,k}(x,\UL) , & j=i,\\
 \chi_{j,k}(x,\UL), & j\neq i.
 \end{cases}
 \end{equation}

We now specialize to specific random matrix ensembles and work out the explicit results using the above general expressions.


\section{Gauss-Wigner ensemble}
\label{SecWig}

\subsection{Correlated case}
We consider $n$-dimensional Hermitian matrices $\bH$ from the distribution
\begin{equation}
\label{WigCorr}
\mathcal{P}(\bH)\propto e^{-\tr \bSg^{-2}\bH^2},
\end{equation}
where, without loss of any generality, we take $\bSg=\diag(\sigma_1,...,\sigma_n)$, with all $\sigma$'s greater than zero. In case $\bSg$ is not already in diagonal form, $\sigma$'s are the corresponding eigenvalues. Similar consideration applies to other correlated ensembles solved ahead.
From Eq.~\eqref{WigCorr} it is evident that the matrix elements of $\bH$ are zero-mean Gaussians having variances as 
\begin{align}
\nonumber
&\langle \bH_{jj}^2\rangle= \frac{\sigma_j^2}{2},~~~~~~~~~~~~~~~~~~~~~~~~~~~~~~~~~~~~~~~~~~\text{(diagonal)} ,\\
&\langle (\text{Re} \bH_{jk})^2\rangle=\langle (\text{Im} \bH_{jk})^2\rangle=\frac{\sigma_j^2\sigma_k^2}{2(\sigma_j^2+\sigma_k^2) },~~~~\text{(off-diagonal)}.
\end{align}
We refer to this case as a correlated Gauss-Wigner ensemble.

The joint eigenvalue density for Eq.~\eqref{WigCorr} can be obtained with the aid of Harish-Chandra-Itzykson-Zuber (HCIZ)-integral~\cite{HC1956,IZ1980} and written down as
\begin{equation}
\label{WigCorrEv}
P(\{\lambda\})=C\frac{\Delta(\{\lambda\})}{\Delta_+(\{\lambda\})}|e^{-\sigma_j^{-2} \lambda_k^2}|_{j,k=1,...,n}.
\end{equation}
Here $\Delta(\{\lambda\})=\prod_{j>k}(\lambda_j-\lambda_k)$ is the Vandermonde determinant, and $\Delta_{+}(\{\lambda\})=\prod_{j>k}(\lambda_j+\lambda_k)$. The eigenvalues lie in $(-\infty,\infty)$. Schur's Pfaffian identity gives~\cite{Schur1911,IOTZ1995,FK2015},
\begin{equation}
\label{SchurPf}
\frac{\Delta(\{\lambda\})}{\Delta_+(\{\lambda\})}=\prod_{j>k}\frac{\lambda_j-\lambda_k}{\lambda_j+\lambda_k}=\begin{cases}
\pf\left[(\lambda_k-\lambda_j)/(\lambda_k+\lambda_j)\right]_{j,k=1,...,n}, & n \text{ even},\\
\pf\begin{bmatrix}\left[(\lambda_k-\lambda_j)/(\lambda_k+\lambda_j)\right]_{j,k=1,...,n} & \left[1\right]_{j=1,...,n} \\
\left[-1\right]_{k=1,...,n} & 0 \end{bmatrix}, & n \text{ odd}.
\end{cases}
\end{equation}
As we can see, this ensemble falls in type II according to our classification, with $w(\lambda)=1$, $f_j(\lambda_k)=e^{-\sigma_j^{-2}\lambda_k^2}$, and $g(\lambda_j,\lambda_k)$ as in Eq.~\eqref{SchurPf}.
The normalization factor in Eq.~\eqref{WigCorrEv} can be obtained using Eq.~\eqref{Ct2} as
\begin{align}
\label{WigCorrNorm}
C^{-1}&=\begin{cases}
n!\pf\left[\pi \sigma_k \sigma_j (\sigma_k^2-\sigma_j^2)/(\sigma_k^2+\sigma_j^2)\right]_{j,k=1,...,n}, & n \text{ even},\\
n!\pf\begin{bmatrix}\left[\pi \sigma_k \sigma_j (\sigma_k^2-\sigma_j^2)/(\sigma_k^2+\sigma_j^2)\right]_{j,k=1,...,n} & \left[\sqrt{\pi}\,\sigma_j\right]_{j=1,...,n} \\
\left[-\sqrt{\pi}\,\sigma_k\right]_{k=1,...,n} & 0 \end{bmatrix}, & n \text{ odd}.
\end{cases}\\
&=n!\,\pi^{n/2}\prod_{i=1}^n \sigma_i \cdot \prod_{j>k}  \frac{\sigma_j^2-\sigma_k^2}{\sigma_j^2+\sigma_k^2}.
\end{align}

\begin{table}
\caption{Gap probabilities: Comparison between analytical and simulation results for correlated Gauss-Wigner ensemble. In each case the $\sigma$ values are from $(\sigma_1,\sigma_2,\sigma_3,\sigma_4,\sigma_5)=(3/4,4/9,1/4,2/7,5/6)$. }
\centering
\begin{tabular}{|c|c|c|c|c|c|c| }
\hline
\multirow{2}{*}{$n$} & \multirow{2}{*}{$r$} & \multirow{2}{*}{$s$} & \multicolumn{2}{|c|}{$E(r,s)$}  & \multicolumn{2}{|c|}{$\widetilde{E}(r,s)$} \\ \cline{4-7}
& & &Analytical  & Simulation &Analytical & Simulation \\
\hline\hline
2 & $-0.5$ & 0.6 & 0.0808 & 0.0804 & 0.3286 & 0.3281 \\
\hline
 3 & 0.1 & 1.4 & 0.0347 & 0.0346 & 0.0017 & 0.0018 \\ 
\hline
 4 & $-\infty$ & $-0.2$ & 0.0111 & 0.0112 & 0.0000 & 0.0000 \\ 
\hline
 4 & $-0.38$ & $0.38$ & $ 0.0034 $ & 0.0031  & 0.0012 & 0.0011 \\ 
\hline
 5 & $-0.7$ & $\infty$ & $ 0.0000 $ & 0.0000  & 0.1541 & 0.1537 \\ 
\hline
\end{tabular}
\label{TabWigCorr}
\end{table}
\begin{figure}[h!]
\centering
 \includegraphics[width=0.75\textwidth]{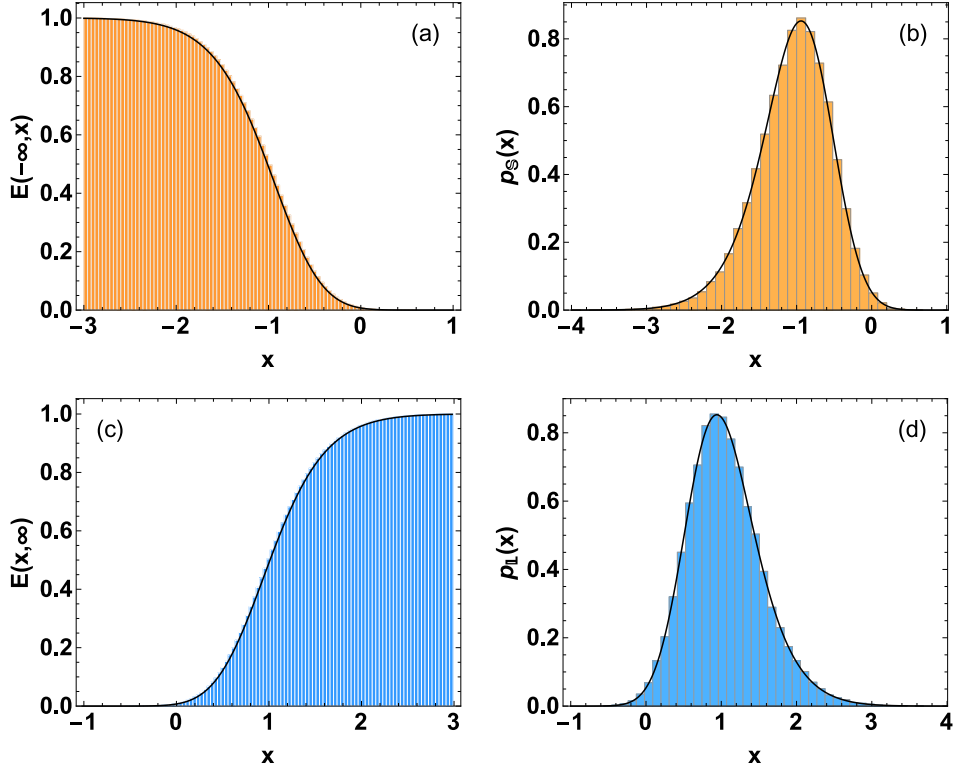}
 \caption{Plots for correlated Gauss-Wigner ensemble with $n=3$, and $(\sigma_1,\sigma_2,\sigma_3)=(3/5,2/3,5/4)$. (a) SF of the smallest eigenvalue (b) PDF of the smallest eigenvalue (c) CDF of the largest eigenvalue (d) PDF of the largest eigenvalue.}
\label{FigWigCorr}
\end{figure}

The gap probability is given by Eq.~\eqref{Ers2} with
\begin{align}
\nonumber
\chi_{j,k}(r,s)&=\int\limits_{(-\infty,r)\cup(s,\infty)} d\lambda \int\limits_{(-\infty,r)\cup(s,\infty)} d\mu\, \frac{\mu-\lambda}{\mu+\lambda}\,e^{-\sigma_j^{-2}\lambda^2}e^{-\sigma_k^{-2}\mu^2} \\
&=\frac{1}{2}\int\limits_{(-\infty,r)\cup(s,\infty)} d\lambda \int\limits_{(-\infty,r)\cup(s,\infty)} d\mu \,\frac{\mu-\lambda}{\mu+\lambda} \left(e^{-\sigma_j^{-2}\lambda^2}e^{-\sigma_k^{-2}\mu^2}-e^{-\sigma_j^{-2}\mu^2}e^{-\sigma_k^{-2}\lambda^2}\right),
\end{align}
and
\begin{align}
\chi_{j,n+1}(r,s)=-\chi_{n+1,j}(r,s)&=(1-\delta_{j,n+1})\int\limits_{(-\infty,r)\cup(s,\infty)} d\lambda \, e^{-\sigma_j^{-2}\lambda^2} \\
&=\frac{\sqrt{\pi}}{2}\sigma_j\left[2+\erf\left(\sigma_j^{-1}r\right)-\erf\left(\sigma_j^{-1}s\right)\right](1-\delta_{j,n+1}),
\end{align}
where $\erf(z)=(2/\sqrt{\pi})\int_0^z dt\, e^{-t^2}$ is the error-function. Similarly $\widetilde{\chi}_{j,k}(r,s)$ gives the double gap probability via Eq.~\eqref{tErs2} with
\begin{align}
\nonumber
\widetilde{\chi}_{j,k}(r,s)&=\int_r^s d\lambda \int_r^s d\mu\,\frac{\mu-\lambda}{\mu+\lambda}\, e^{-\sigma_j^{-2}\lambda^2}e^{-\sigma_k^{-2}\mu^2} \\
&=\frac{1}{2}\int_r^s d\lambda \int_r^s d\mu \, \frac{\mu-\lambda}{\mu+\lambda}\left(e^{-\sigma_j^{-2}\lambda^2}e^{-\sigma_k^{-2}\mu^2}-e^{-\sigma_j^{-2}\mu^2}e^{-\sigma_k^{-2}\lambda^2}\right),
\end{align}
and
\begin{align}
\nonumber
\widetilde{\chi}_{j,n+1}(r,s)=-\chi_{n+1,j}(r,s)&=(1-\delta_{j,n+1})\int_r^s d\lambda \, e^{-\sigma_j^{-2}\lambda^2} \\
&=\frac{\sqrt{\pi}}{2}\sigma_j\left[\erf\left(\sigma_j^{-1}s\right)-\erf\left(\sigma_j^{-1}r\right)\right](1-\delta_{j,n+1}).
\end{align}
The expressions for density of the extreme eigenvalues follow from Eqs.~\eqref{pS2} and~\eqref{pL2}. 

In Table~\ref{TabWigCorr} we compare the results for gap probabilities for various choice of parameters. The simulation results have been calculated using 50000 matrix realizations, as has been also done for other ensembles ahead. Fig.~\ref{FigWigCorr} shows the plots of the distribution functions and densities obtained from analytical expressions (solid lines) as well as from Monte-Carlo simulations (histograms). We find excellent agreement in all cases. We note that there is a symmetry $\chi(-\infty,-x)=\chi(x,\infty)$ in this case. It leads to $E(-\infty,-x)=E(x,\infty)$ and $p_S(-x)=p_L(x)$ which are clearly observed in the plots. These correspondences are consequence of the symmetry exhibited by the matrix model~\eqref{WigCorr}. 


\subsection{Uncorrelated case}

We consider here the case $\bSg=\1$ in~\eqref{WigCorr}, and hence obtain
\begin{equation}
\label{WigUn}
\mathcal{P}(\bH)\propto e^{-\tr \bH^2},
\end{equation}
which is the density for the classical Gaussian Unitary Ensemble (GUE)~\cite{Mehta2004,Forrester2010}. In this case the variance of diagonal elements of $\bH$ is 1/2, while that of the real and imaginary parts of off-diagonal elements is 1/4. The results for the present case can be obtained from the preceding section using a limiting procedure to set all $\sigma$ equal to 1. However, it is more straightforward to start with Eq.~\eqref{WigUn}, which gives the following result for the joint density of eigenvalues:
\begin{equation}
\label{WigUnEv}
P(\lambda_1,...,\lambda_N)=C\Delta^2(\{\lambda\}) \prod_{j=1}^n e^{-\lambda_j^2}.
\end{equation}
It is interesting to see how the product of a Pfaffian and determinant gives rise to the product of two determinants for $\bSg=\1$. Comparing Eq.~\eqref{WigUnEv} with our type I ensemble form, Eq.~\eqref{Type1}, we find that $f_j(\lambda_k)=g_j(\lambda_k)=\lambda_k^{j-1}$, and weight function $w(\lambda)=e^{-\lambda^2}$.
The normalization factor in Eq.~\eqref{WigUnEv} is well known~\cite{Mehta2004,Forrester2010}:
\begin{align}
\nonumber
C^{-1}&=n!\,\Big|\frac{1+(-1)^{j+k}}{2}\,\Gamma\Big(\frac{j+k-1}{2}\Big)\Big|_{j,k=1,...,n}\\
&=\frac{\pi^{n/2}}{2^{n(n-1)/2}}\prod_{j=1}^n \Gamma(j+1)=\frac{\pi^{n/2}}{2^{n(n-1)/2}}G(n+2),
\end{align}
where $\Gamma(z)=\int_0^\infty dt\,t^{z-1} e^{-t}$ is the Gamma function, and $G(n)$  is the Barnes $G$-function which is defined for a positive integer $n$ as $G(n+1)=\prod_{j=1}^{n} \Gamma(j)$.

Gaussian ensembles are probably the most studied random matrix ensembles, along with the Laguerre-Wishart ones~\cite{Mehta2004,Forrester2010}. For large $n$, the distributions of extreme eigenvalues for Gaussian ensembles, including the complex case above, were derived by Tracy and Widom in~\cite{TW1994,TW1996}. Further progress has been made in~\cite{DM2006,DM2008,MV2009} where large deviations of the extreme eigenvalues have been explored in detail. We are interested here in finite $n$ exact results.

The expression for gap probability is given by Eq.~\eqref{Ers1} with 
\begin{equation}
\label{WigUnChi}
\chi_{j,k}(r,s)
=\begin{cases}
\frac{1}{2}\left[(1+(-1)^{j+k})\,\Gamma(\frac{j+k-1}{2})+(-1)^{j+k}\Big(\Gamma(\frac{j+k-1}{2},r^2)-\Gamma(\frac{j+k-1}{2},s^2)\Big)\right], & r<s\le0,\\
\frac{1}{2}\left[(-1)^{j+k}\Gamma(\frac{j+k-1}{2},r^2)+\Gamma(\frac{j+k-1}{2},s^2)\right], & r<0<s,\\
\frac{1}{2}\left[(1+(-1)^{j+k})\,\Gamma(\frac{j+k-1}{2})-\Gamma(\frac{j+k-1}{2},r^2)+\Gamma(\frac{j+k-1}{2},s^2)\right], & 0\le r<s.
\end{cases}
\end{equation} 
Note that when $j+k$ is odd, all the three expressions above become identical. The expression for double gap probability $\widetilde{E}(r,s)$ is given by Eq.~\eqref{tErs1} with
\begin{equation}
\widetilde{\chi}_{j,k}(r,s)
=\begin{cases}
\frac{1}{2}(-1)^{j+k}\left[-\Gamma(\frac{j+k-1}{2},r^2)+\Gamma(\frac{j+k-1}{2},s^2)\right], & r<s\le0,\\
\frac{1}{2}\left[(1+(-1)^{j+k})\,\Gamma(\frac{j+k-1}{2})-(-1)^{j+k}\Gamma(\frac{j+k-1}{2},r^2)-\Gamma(\frac{j+k-1}{2},s^2)\right], & r<0<s,\\
\frac{1}{2}\left[\Gamma(\frac{j+k-1}{2},r^2)-\Gamma(\frac{j+k-1}{2},s^2)\right], & 0\le r<s.
\end{cases}
\end{equation} 
 These reduce to the following for $r=-\infty$ or $s=\infty$:
 \begin{equation}
\chi_{j,k}(-\infty,x)=\widetilde{\chi}_{j,k}(x,\infty)
=\begin{cases}
\frac{1}{2}\left[(1+(-1)^{j+k})\,\Gamma(\frac{j+k-1}{2})-(-1)^{j+k}\Gamma(\frac{j+k-1}{2},x^2)\right], & x\le0,\\
\frac{1}{2}\Gamma(\frac{j+k-1}{2},x^2), & x>0,
\end{cases}
 \end{equation}
 \begin{equation}
\chi_{j,k}(x,\infty)=\widetilde{\chi}_{j,k}(-\infty,x)
=\begin{cases}
\frac{1}{2}\left[(-1)^{j+k}\Gamma(\frac{j+k-1}{2},x^2)\right], & x\le 0,\\
\frac{1}{2}\left[(1+(-1)^{j+k})\,\Gamma(\frac{j+k-1}{2})-\Gamma(\frac{j+k-1}{2},x^2)\right], & x>0.
\end{cases}
 \end{equation}

\begin{table}
\caption{Gap probabilities: Comparison between analytical and simulation results for uncorrelated Gauss-Wigner ensemble. }
\centering
\begin{tabular}{|c|c|c|c|c|c|c| }
\hline
\multirow{2}{*}{$n$} & \multirow{2}{*}{$r$} & \multirow{2}{*}{$s$} & \multicolumn{2}{|c|}{$E(r,s)$}  & \multicolumn{2}{|c|}{$\widetilde{E}(r,s)$} \\ \cline{4-7}
& & &Analytical  & Simulation &Analytical & Simulation \\
\hline\hline
2 & $-2$ & $-0.2$ & 0.1814 & 0.1811 & 0.0418 & 0.0421 \\
\hline
3 & 0.05 & $\infty$ & 0.0074 & 0.0075 & 0.0042 & 0.0041 \\ 
\hline
3 & $-\infty$ & 0 & 0.0056 & 0.0056 & 0.0056 & 0.0053 \\ 
\hline
4 & $-1.2$ & 1.2 & 0.0018 & 0.0019 & 0.0073 & 0.0073 \\ 
\hline
5 & $-1.4$ & 4 & 0.0000 & 0.0000 & 0.0589 & 0.0587 \\ 
\hline
\end{tabular}
\label{TabWigUn}
\end{table}

The density of smallest eigenvalue is given by Eq.~\eqref{pS1} with
\begin{equation}
\phi_{j,k}^{(i)}(x)
=\begin{cases}
e^{-x^2}x^{j+k-2}, & j=i,\\
 \chi_{j,k}(-\infty,x), & j\neq i.
 \end{cases}
 \end{equation}
and that of the largest eigenvalue by Eq.~\eqref{pL1} with
\begin{equation}
\psi_{j,k}^{(i)}(x)
=\begin{cases}
e^{-x^2}x^{j+k-2}, & j=i,\\
 \chi_{j,k}(x,\infty), & j\neq i.
 \end{cases}
 \end{equation}

\begin{figure}[ht!]
\centering
 \includegraphics[width=0.75\textwidth]{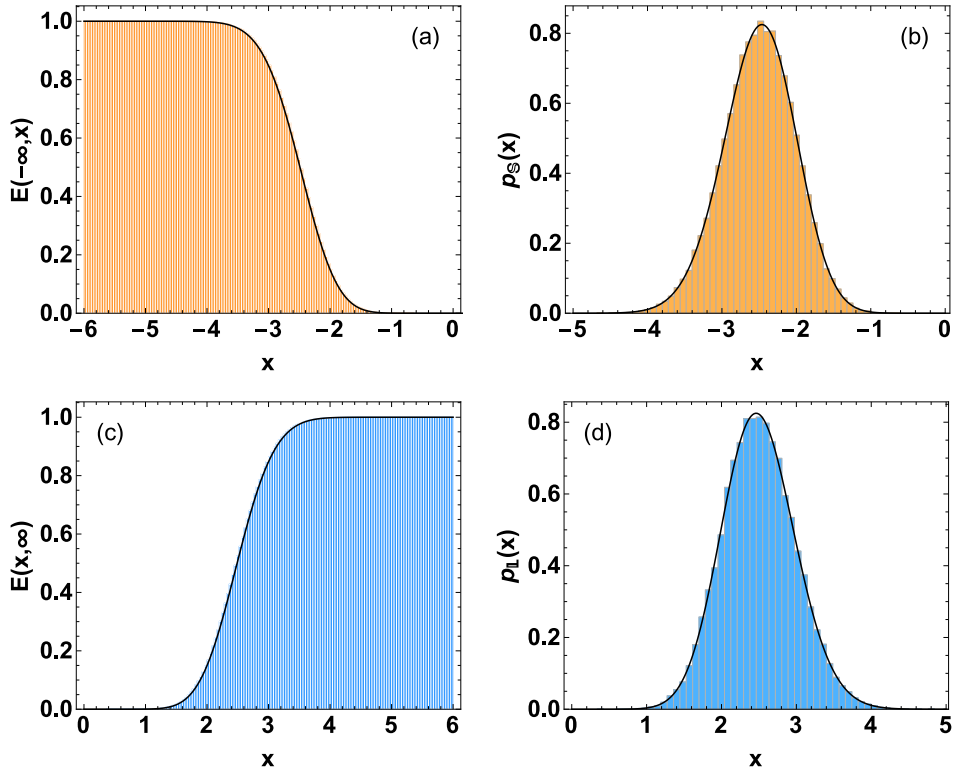}
 \caption{Plots for uncorrelated Gauss-Wigner ensemble with $n=6$. (a) SF of the smallest eigenvalue (b) PDF of the smallest eigenvalue (c) CDF of the largest eigenvalue (d) PDF of the largest eigenvalue.}
\label{FigWigUn}
\end{figure}

In Table~\ref{TabWigUn} we compare the results for gap probabilities for various combination of parameters. Fig.~\ref{FigWigUn} shows the analytical distributions and densities, as well as histograms obtained from Monte-Carlo simulations. We again find the symmetries $E(-\infty,-x)=E(x,\infty)$ and  $p_S(-x)=p_L(x)$ manifest in the plots. We note, however, in this case that $\chi_{j,k}(-\infty,-x)=(-1)^{j+k}\chi_{j,k}(x,\infty)$ and the aforementioned symmetries follow from the determinanatal expression~\eqref{Ers1}. The underlying reason for these is again the matrix probability density.


\section{Laguerre-Wishart ensemble}
\label{SecWis}

\subsection{Correlated case}

The correlated Laguerre-Wishart ensemble is described by the density
\begin{equation}
\label{WisCorr}
\mathcal{P}(\bH)\propto |\bH|^\alpha e^{-\tr \bSg^{-1}\bH},
\end{equation}
for positive-definite (or more generally nonnegative-definite) Hermitian matrices $\bH$ with $\alpha> -1$ and $\bSg=\diag(\sigma_1,...,\sigma_n)$, $\sigma_j>0$. For nonnegative integer values of $\alpha$ this ensemble can be realized as follows. Consider $n\times m$-dimensional ($n\le m$) complex Gaussian matrices $\bG$ from the distribution
\begin{equation}
\label{GinibreCorr}
\mathcal{P}_G(\bG)\propto e^{-\tr \bSg^{-1}\bG\bG^\dag},
\end{equation}
then $n$-dimensional positive-definite Hermitian matrices (Wishart matrices)
\begin{equation}
\label{GGd}
\bH=\bG\bG^\dag
\end{equation}
constitute the correlated Laguerre-Wishart ensemble with the associated density given by Eq.~\eqref{WisCorr}, and the parameter $\alpha=m-n$. We underline that the $m\times m$ matrices $\bG^\dag\bG$ have $m-n$ generic zero eigenvalues, while the $n$ nonzero eigenvalues are same as those of $\bG\bG^\dag$. The parameter $m$ is referred to as the degree of freedom of the Wishart distribution. 
Equivalent to the above construction, if $n\times m$-dimensional ($n\le m$) complex (Ginibre) matrices $\widetilde{\bG}$ are taken from
\begin{equation}
\label{GinibreUn}
\mathcal{P}_{\widetilde{G}}(\widetilde{\bG})\propto e^{-\tr \widetilde{\bG}\widetilde{\bG}^\dag},
\end{equation}
then $\bH$ of Eq.~\eqref{WisCorr} with $\alpha=m-n$ can be generated as
\begin{equation}
\label{SGGdSd}
\bH=\bSg^{1/2}\widetilde{\bG}\widetilde{\bG}^\dag\bSg^{1/2}.
\end{equation}

The joint eigenvalue density for the correlated Laguerre-Wishart ensemble can be obtained with the aid of HCIZ-integral, and written in a biorthogonal form as
\begin{equation}
\label{WisCorrEv}
P(\lambda_1,...,\lambda_n)=C \Delta(\{\lambda\})\prod_{l=1}^n \lambda_l^{\alpha}\cdot|e^{-\sigma_j^{-1} \lambda_k}|_{j,k=1,...,n},
\end{equation}
with $0\leq\lambda_j<\infty$. This JPDF falls in type I as per our classification with $w(\lambda)=\lambda^\alpha$, $f_j(\lambda_k)=\lambda_k^{j-1}$ and $g_j(\lambda_k)=e^{-\sigma_j^{-1}\lambda_k}$. The partition function for Eq.~\eqref{WisCorrEv} is
\begin{align}
\nonumber
C^{-1}&=n!\,|\sigma_k^{j+\alpha}\,\Gamma(j+\alpha)|_{j,k=1,...,n}\\
&=n!\,\Delta(\{\sigma\})\prod_{j=1}^n\sigma_j^{\alpha+1}\,\Gamma(j+\alpha).
\end{align}
Eq.~\eqref{GGd} or~\eqref{SGGdSd} provides a way to easily generate matrices and hence the eigenvalues distributed according to Eq.~\eqref{WisCorrEv} for non-negative integer $\alpha$ values, while for any real $\alpha\, (\,>-1)$ the eigenvalues can be generated with the aid of Monte-Carlo simulation based on Dyson's log-gas picture~\cite{GPPS2003,Kumar2014}, as already indicated in Section~\ref{SecGen}.

Laguerre-Wishart ensemble has been explored extensively because of their crucial role in the field of multivariate statistics~\cite{Anderson2003,Muirhead2005}, and explicit appearance in problems related to time series~\cite{Gnanadesikan1997,PGRAGS2002,VP2010} and multiple-channel telecommunication~\cite{CTKV2002,Burel2002,ZNYZY2008,NZYY2008,ZCW2009,FG1998,Telatar1999,KP2010a}. As a consequence several results concerning extreme eigenvalues are available. Some of the most recent results being due to Wirtz {\it et al.}~\cite{WG2013,WG2014,WKG2015,AGKWW,WAGKW}. Our results for gap probabilities and PDF of extreme eigenvalues possess form as in~\cite{WG2013,WG2014}, however the kernels involved have relatively simpler and closed structure. 

The gap probability in this case is given by Eq.~\eqref{Ers1} with
\begin{equation}
\chi_{j,k}(r,s)=\sigma_k^{j+\alpha}[\gamma(j+\alpha,\sigma_k^{-1}r)+\Gamma(j+\alpha,\sigma_k^{-1}s)],
\end{equation}
where $\gamma(a,r)=\int_0^r dz\, z^{a-1} e^{-z}$ and $\Gamma(a,r)=\int_s^\infty dz\, z^{a-1} e^{-z}$ represent the lower-incomplete gamma function and the upper-incomplete gamma function, respectively. 
Also, for the double gap probability, as in Eq.~\eqref{tErs1}, we have
\begin{equation}
\widetilde{\chi}_{j,k}(r,s)=\sigma_k^{j+\alpha}[\Gamma(j+\alpha,\sigma_k^{-1}r)-\Gamma(j+\alpha,\sigma_k^{-1}s)].
\end{equation}

These simplify to the following for $r=0$ or $s=\infty$:
\begin{equation}
\chi_{j,k}(0,x)=\widetilde{\chi}(x,\infty)=\sigma_k^{j+\alpha}\Gamma(j+\alpha,\sigma_k^{-1}x),
\end{equation}
\begin{equation}
\chi_{j,k}(x,\infty)=\widetilde{\chi}(0,x)=\sigma_k^{j+\alpha}\gamma(j+\alpha,\sigma_k^{-1}x).
\end{equation}

\begin{table}
\caption{Gap probabilities: Comparison between analytical and simulation results for correlated Laguerre-Wishart ensemble. The $\sigma$ values are from $(\sigma_1,...,\sigma_6)=(2,3/5,7/3,2/5,4/9,1/2)$. The $m$ values are indicated when matrix construction is possible using Eq.~\eqref{GGd} or~\eqref{SGGdSd}. }
\centering
\begin{tabular}{|c|c|c|c|c|c|c|c| c| }
\hline
\multirow{2}{*}{$n$} & \multirow{2}{*}{$\alpha$}  & \multirow{2}{*}{$m$} & \multirow{2}{*}{$r$} & \multirow{2}{*}{$s$} & \multicolumn{2}{|c|}{$E(r,s)$}  & \multicolumn{2}{|c|}{$\widetilde{E}(r,s)$} \\ \cline{6-9}
& & & & & Analytical  & Simulation &Analytical & Simulation \\
\hline\hline
2 & 1.25 & $-$ & 0.63 & ~~5~~ & 0.1789 & 0.1786 & 0.1913 & 0.1926 \\
\hline
~~2~~ & 3 & ~~5~~ & ~3.1~ & $\infty$ & 0.0046 & 0.0047 & 0.1753 & 0.1747\\ 
\hline
3 & 2 & 5 & 0 & 5 & 0.0013 & 0.0013 & 0.0006 & 0.0007\\ 
\hline
4 & 4 & 8 & 2.5 & 20 & 0.0020 & 0.0020 & 0.0103 & 0.0105 \\ 
\hline
5 & $-0.3$ & $-$ & 4 & 24 & 0.0051 & 0.0053 & 0.0000 & 0.0000\\ 
\hline
6 & 0 & 6 & 0.3 & 28 & 0.0000 & 0.0000 & 0.0523 & 0.0521\\ 
\hline
\end{tabular}
\label{TabWisCorr}
\end{table}
\begin{figure}[t!]
\centering
 \includegraphics[width=0.75\textwidth]{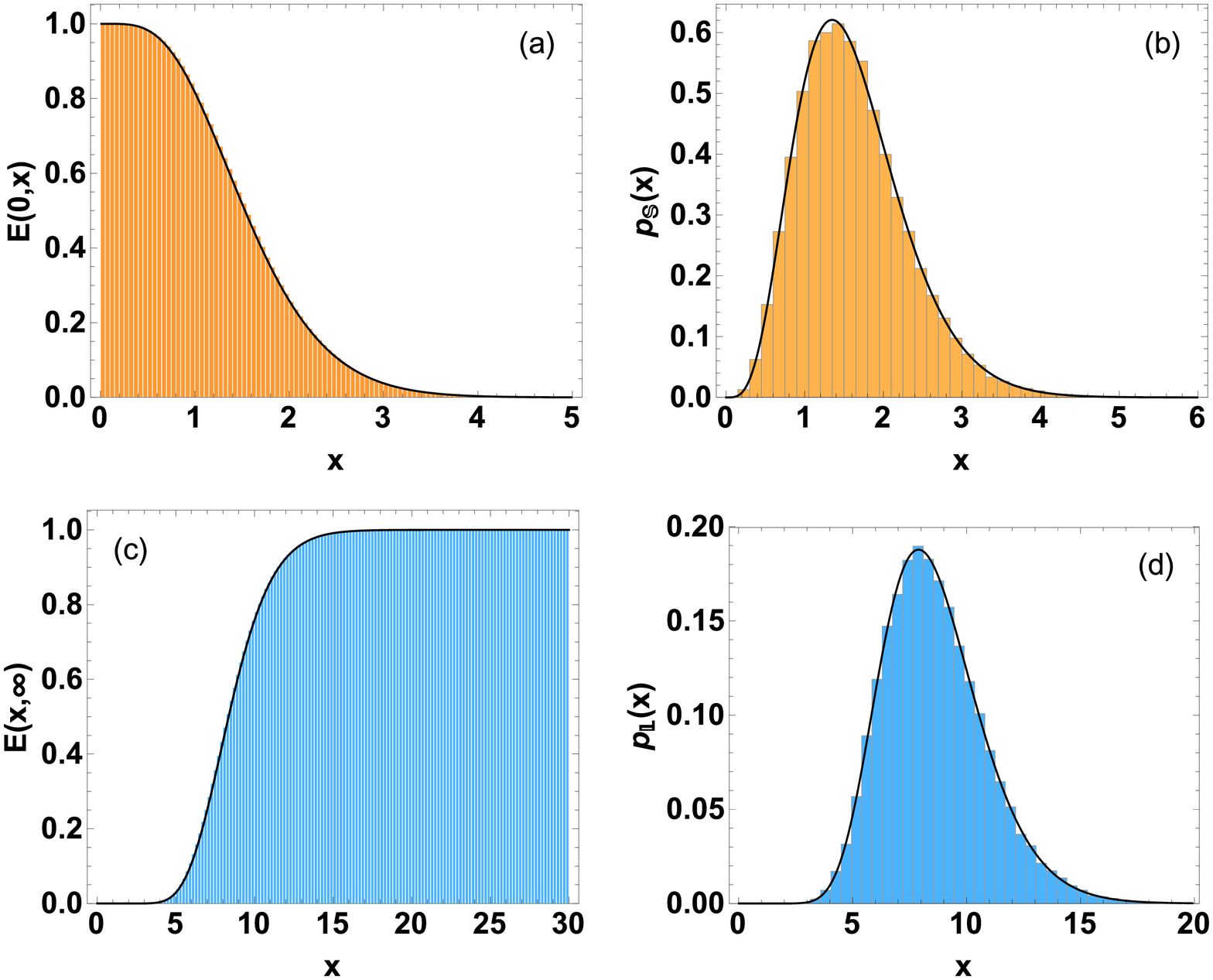}
 \caption{Plots for correlated Laguerre-Wishart ensemble with $n=3, m=7$, and $\bSg=(1/2,3/4,4/5)$. (a) SF of the smallest eigenvalue (b) PDF of the smallest eigenvalue (c) CDF of the largest eigenvalue (d) PDF of the largest eigenvalue.}
\label{FigWisCorr}
\end{figure}

The expressions for the smallest eigenvalue and the largest eigenvalue are given respectively by Eqs.~\eqref{pS1} and~\eqref{pL1} with
\begin{equation}
\phi_{j,k}^{(i)}(x)
=\begin{cases}
x^{j+\alpha-1}\,e^{-\sigma_k^{-1}x}, & j=i,\\
\chi(0,x), & j\neq i,
 \end{cases}
 \end{equation}
and
\begin{equation}
\psi_{j,k}^{(i)}(x)
=\begin{cases}
x^{j+\alpha-1}\,e^{-\sigma_k^{-1}x}, & j=i,\\
\chi(x,\infty), & j\neq i.
 \end{cases}
 \end{equation}

In Table~\ref{TabWisCorr} we list the gap probabilities for several choices of parameters. Fig.~\ref{FigWisCorr} displays the plots of distribution functions and probability density functions obtained from analytical expressions, which are validated by the histograms obtained from Monte-Carlo simulation.

A particularly interesting scenario occurs when $\alpha=0$, i.e., $m=n$. In this case the gap probability $E(0,x)$ or $\widetilde{E}(x,\infty)$ and the density of the smallest eigenvalue possess remarkably simple expression. These can be obtained conveniently if we start from the JPDF in Eq.~\eqref{WisCorr} and then implement the translation $\mu_j=\lambda_j-x$. Upon a little simplification, the gap probability is obtained as
\begin{equation}
\label{Emeqn}
E(0,x)=\widetilde{E}(x,\infty)=\exp\left(-x\sum_{j=1}^n \sigma_j^{-1} \right).
\end{equation}
The density of the smallest eigenvalue then follows immediately as
\begin{equation}
\label{pSmeqn}
p_\mathds{S}(x)=\left(\sum_{j=1}^n \sigma_j^{-1} \right)\exp\left(-x\sum_{j=1}^n \sigma_j^{-1} \right).
\end{equation}


\subsection{Uncorrelated case}

We now deal with the uncorrelated Laguerre-Wishart case which involves positive-definite Hermitian matrices $\bH$ from the density
\begin{equation}
\label{WisUn}
\mathcal{P}(\bH)\propto |\bH|^{\alpha}e^{-\tr \bH},
\end{equation}
with $\alpha>-1$. Again, for nonnegative integer values of $\alpha \,(=m-n)$ the matrices from the above density can be generated as $\bH=\widetilde{\bG}\widetilde{\bG}^\dag$, where $\widetilde{\bG}$ are $n\times m$ dimensional ($n\le m$) complex matrices taken from the density~\eqref{GinibreUn}.

Equation~\eqref{WisUn} is recognized as the standard Laguerre Unitary Ensemble of random matrices with the corresponding joint eigenvalue density~\cite{Mehta2004,Forrester2010}:
\begin{equation}
P(\lambda_1,...,\lambda_N)=C\Delta^2(\{\lambda\}) \prod_{j=1}^n \lambda_j^{\alpha}e^{-\lambda_j}.
\end{equation}
This is ensemble of type I according to our classification. When compared with Eq.~\eqref{Type1} we find that $f_j(\lambda_k)=g_j(\lambda_k)=\lambda_k^{j-1}$, and $w(\lambda)=\lambda^{\alpha}e^{-\lambda}$.
The normalization factor is well known~\cite{Mehta2004,Forrester2010}:
\begin{align}
\nonumber
C^{-1}&=n!\,|\Gamma(j+k+\alpha-1)|_{j,k=1,...,n}\\
&=\prod_{j=1}^n \Gamma(j+1)\Gamma(j+\alpha).
\end{align}

The expression for gap probability is given by Eq.~\eqref{Ers2} with the kernel
\begin{equation}
\chi_{j,k}(r,s)=\gamma(j+k+\alpha-1,r)+\Gamma(j+k+\alpha-1,s).
\end{equation} 
Similarly, the double gap probability is given by Eq.~\eqref{tErs2} with
\begin{equation}
\widetilde{\chi}_{j,k}(r,s)=\Gamma(j+k+\alpha-1,r)-\Gamma(j+k+\alpha-1,s).
\end{equation} 
For $r=0$ or $s=\infty$ these yield
\begin{align}
\chi_{j,k}(0,x)=\widetilde{\chi}_{j,k}(x,\infty)=\Gamma(j+k+\alpha-1,x),
\end{align}
\begin{align}
\chi_{j,k}(x,\infty)=\widetilde{\chi}_{j,k}(0,x)=\gamma(j+k+\alpha-1,x).
\end{align}

\begin{table}
\caption{Gap probabilities: Comparison between analytical and simulation results for uncorrelated Laguerre-Wishart ensemble. The $m$ values are indicated in the cases when matrix construction is possible using $\widetilde{\bG}\widetilde{\bG}^\dag$ where $\widetilde{\bG}$ is from Eq.~\eqref{GinibreUn}. }
\centering
\begin{tabular}{|c|c|c|c|c|c|c|c|c|}
\hline
\multirow{2}{*}{$n$} & \multirow{2}{*}{$\alpha$} & \multirow{2}{*}{$m$} & \multirow{2}{*}{$r$} & \multirow{2}{*}{$s$} & \multicolumn{2}{|c|}{$E(r,s)$}  & \multicolumn{2}{|c|}{$\widetilde{E}(r,s)$} \\ \cline{6-9}
& & &  & &Analytical  & Simulation &Analytical & Simulation \\
\hline\hline
~~2~~ & ~~2~~ & ~~4~~ & ~~1~~ & ~~5~~ & 0.1221 & 0.1224 & 0.2298 & 0.2300 \\
\hline
3 & $-0.25$ & $-$ & 0.9 & 9 & 0.0085 & 0.0088 & 0.0273 & 0.0269 \\ 
\hline
4 & 0 & 4 & 1.5 & 14 & 0.0001 & 0.0001 & 0.0021 & 0.0020\\
\hline
5 & 2.6 & $-$ & 0 & 15 & 0.0000 & 0.0000 & 0.2554 & 0.2546 \\
\hline
6 & 3 & 9 & 18 & $\infty$ & 0.1643 & 0.1640 & 0.0000 & 0.0000\\ 
\hline
\end{tabular}
\label{TabWisUn}
\end{table}
\begin{figure}[ht!]
\centering
 \includegraphics[width=0.75\textwidth]{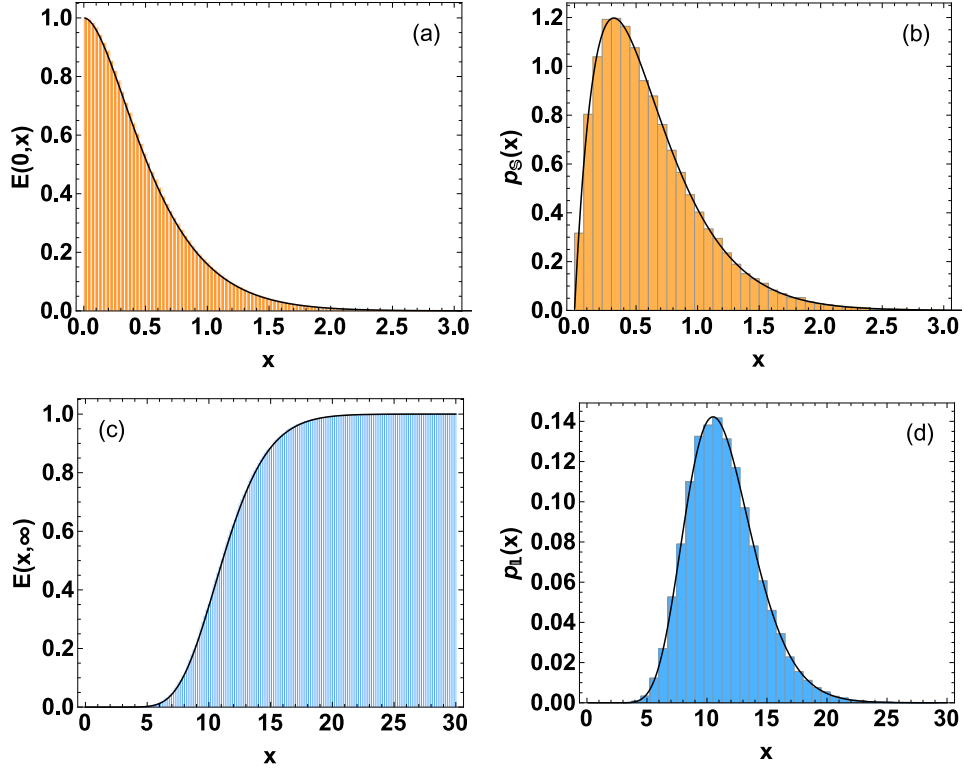}
\caption{Plots for uncorrelated Laguerre-Wishart ensemble with $n=4, m=5$. (a) SF of the smallest eigenvalue (b) PDF of the smallest eigenvalue (c) CDF of the largest eigenvalue (d) PDF of the largest eigenvalue.}
\label{FigWisUn}
\end{figure}

The expressions for densities of the extreme eigenvalues are given by Eqs.~\eqref{pS1} and~\eqref{pL1} with
\begin{equation}
\phi_{j,k}^{(i)}(x)
=\begin{cases}
x^{j+k+\alpha-2}\,e^{-x}, & j=i,\\
\chi(0,x), & j\neq i,
 \end{cases}
 \end{equation}
and
\begin{equation}
\psi_{j,k}^{(i)}(x)
=\begin{cases}
x^{j+k+\alpha-2}\,e^{-x}, & j=i,\\
\chi(x,\infty), & j\neq i.
 \end{cases}
 \end{equation}
 
 Table~\ref{TabWisUn} displays the results for gap probabilities with comparison between analytical result and numerical simulation. Fig.~\ref{FigWisUn} shows the plots for survival function for the smallest eigenvalue, cumulative distribution function for the largest eigenvalue, and the corresponding densities.
  
Following the correlated case, Eqs.~\eqref{Emeqn} and \eqref{pSmeqn}, for $\alpha=0\, (m=n)$ the gap probability $E(0,x)$ or $\widetilde{E}(x,\infty)$ and the density of the smallest eigenvalue have the following simple expressions~\cite{Edelman1991}:
\begin{equation}
E(0,x)=\widetilde{E}(x,\infty)=e^{-n x},
\end{equation}
\begin{equation}
p_\mathds{S}(x)=n e^{-nx} .
\end{equation}


\section{Cauchy-Lorentz ensemble: Variant I}
\label{SecCL1}

\subsection{Correlated case}
We refer to the ensemble of $n$-dimensional Hermitian matrices $\bH$ from the density
\begin{equation}
\label{CL1Corr}
\mathcal{P}(\bH)\propto|\1+\bSg^{-2}\bH^2|^{-\kappa},
\end{equation}
as the correlated Cauchy-Lorentz (variant I) ensemble. Here $\bSg=\diag(\sigma_1,...,\sigma_n), \sigma_j>0$, and $\kappa>n-1/2$ for convergence.
Using unitary group integral result of~\cite{Orlov2004} we arrive at the joint eigenvalue density
\begin{equation}
P(\{\lambda\})=C\frac{\Delta(\{\lambda\})}{\Delta_+(\{\lambda\})}|(1+\sigma_j^{-2} \lambda_k^2)^{-\kappa+n-1}|_{j,k=1,...,n},
\end{equation}
with $-\infty < \lambda_j<\infty$. Similar to the correlated Gauss-Wigner case, the above JPDF can be rewritten as product of a determinant  and a Pfaffian.
Therefore, the normalization factor is provided by a Pfaffian expression as
\begin{equation}
\nonumber
C^{-1}=\pf[h_{jk}]_{j,k=1,..,N},
\end{equation}
where $N$ is as defined in Eq.~\eqref{DefN}. Also, for even $n$ we have
\begin{align}
\nonumber
h_{j,k}=\frac{\pi\,\Gamma^2(\kappa-n+1/2)}{2\,\Gamma^2(\kappa-n+1)}\Big[\sigma_j^{2(\kappa-n+1)}\sigma_k^{-2(\kappa-n)}\,_2F_1\Big(2(\kappa-n)+1,\kappa-n+3/2;2(\kappa-n+1);1-\sigma_k^{-2}\sigma_j^2\Big)\\
-\sigma_k^{2(\kappa-n+1)}\sigma_j^{-2(\kappa-n)}\,_2F_1\Big(2(\kappa-n)+1,\kappa-n+3/2;2(\kappa-n+1);1-\sigma_j^{-2}\sigma_k^2\Big)\Big].
\end{align}
and when $n$ is odd, additionally we have
\begin{align}
\nonumber
h_{j,n+1}=-h_{n+1,j}&=(1-\delta_{j,n+1})\int_{-\infty}^\infty d\lambda \,(1+\sigma_j^{-2} \lambda^2)^{-\kappa+n-1}\\
&=\sqrt{\pi}\sigma_j\frac{\Gamma(\kappa-n+1/2)}{\Gamma(\kappa-n+1)}(1-\delta_{j,n+1}).
\end{align}
In the above expression for $h_{j,k}$, $_2F_1(a,b;c;z)=\frac{\Gamma(c)}{\Gamma(a)\Gamma(b)}\sum_{k=0}^\infty \frac{\Gamma(a+k)\Gamma(b+k)}{\Gamma(c+k)}\frac{z^k}{k!}$ is the Gauss hypergeometric function. For $\kappa=n$, the partition function $C^{-1}$ simplifies to
\begin{align}
C^{-1}&=\begin{cases}
n!\pf\left[\pi^2\sigma_k\sigma_j(\sigma_k-\sigma_j)/(\sigma_k+\sigma_j)\right]_{j,k=1,...,n}, & n \text{ even},\\
n!\pf\begin{bmatrix}\left[\pi^2\sigma_k\sigma_j(\sigma_k-\sigma_j)/(\sigma_k+\sigma_j)\right]_{j,k=1,...,n} & \left[\pi\sigma_j\right]_{j=1,...,n} \\
\left[-\pi\sigma_k\right]_{k=1,...,n} & 0 \end{bmatrix}, & n \text{ odd}.
\end{cases}\\
&=n!\,\pi^n\prod_{i=1}^n \sigma_i\,\cdot \prod_{j>k}\frac{\sigma_j-\sigma_k}{\sigma_j+\sigma_k}.
\end{align}
\begin{table}
\caption{Gap probabilities: Comparison between analytical and simulation results for correlated Cauchy-Lorentz (variant I) ensemble. The $\sigma$ values are from $(\sigma_1,...,\sigma_6)=(5/6,7/8,16/11,2,3/10)$. }
\centering
\begin{tabular}{|c|c|c|c|c|c|c|c| }
\hline
\multirow{2}{*}{$n$} & \multirow{2}{*}{$\kappa$} & \multirow{2}{*}{$r$} & \multirow{2}{*}{$s$} & \multicolumn{2}{|c|}{$E(r,s)$}  & \multicolumn{2}{|c|}{$\widetilde{E}(r,s)$} \\ \cline{5-8}
& & & &Analytical  & Simulation &Analytical & Simulation \\
\hline\hline
~2~ & 3 & $-3$ & 0 & 0.1251 & 0.1242 & 0.1078  & 0.1073 \\
\hline
3 & 3.2 & $-0.5$ & 1.2 & 0.0471 & 0.0492 & 0.0116 & 0.0118 \\ 
\hline
3 & 4 & $-0.6$ & 0.7 & 0.0356 & 0.0354 & 0.0195 & 0.0198 \\ 
\hline
4 & 4.5 & $2.6$ & 16 & 0.6117 & 0.6108 & 0.0000 & 0.0000 \\ 
\hline
5 & 5 & $-9.5$ & 9.5& 0.0000 & 0.0000 & 0.6423 & 0.6440 \\ 
\hline
\end{tabular}
\label{TabCL1Corr}
\end{table}
\begin{figure}[ht!]
\centering
 \includegraphics[width=0.75\textwidth]{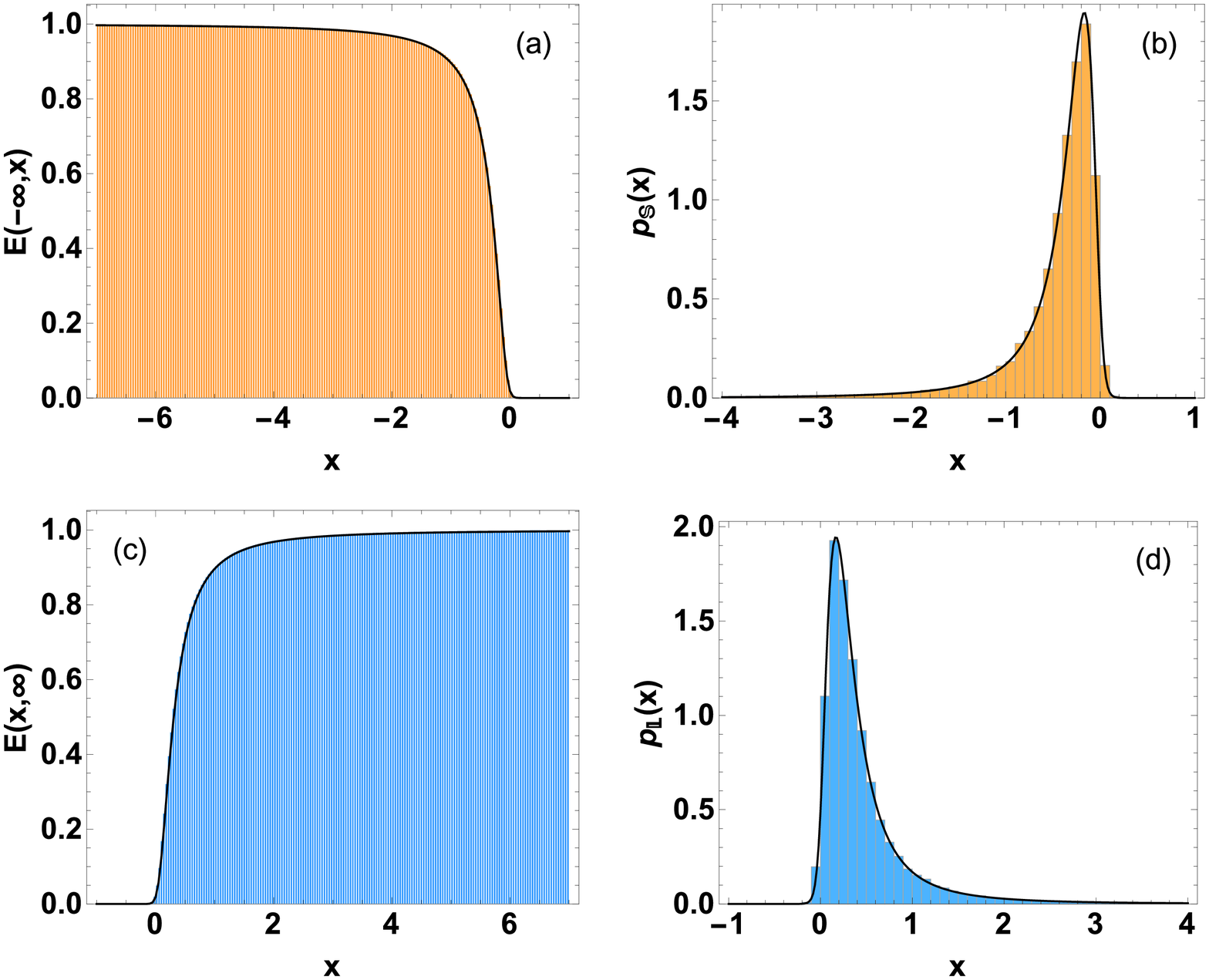}
 \caption{Plots for correlated Cauchy-Lorentz (Variant I) ensemble with $n=3, \kappa=3.4$, and $\bSg=(1/7,3/8,2/9)$. (a) SF of the smallest eigenvalue (b) PDF of the smallest eigenvalue (c) CDF of the largest eigenvalue (d) PDF of the largest eigenvalue.}
\label{FigCL1Corr}
\end{figure}

The gap probability $E(r,s)$ is given by Eq.~\eqref{Ers2} with
\begin{align}
\nonumber
&\chi_{j,k}(r,s)=\int\limits_{(-\infty,r)\cup(s,\infty)} d\lambda \int\limits_{(-\infty,r)\cup(s,\infty)} d\mu\,\frac{\mu-\lambda}{\mu+\lambda} \left[(1+\sigma_j^{-2} \lambda^2)^{-\kappa+n-1}(1+\sigma_k^{-2} \mu^2)^{-\kappa+n-1}\right] \\
&=\frac{1}{2}\!\!\!\!\int\limits_{(-\infty,r)\cup(s,\infty)} \!\!\!\!d\lambda \!\! \int\limits_{(-\infty,r)\cup(s,\infty)}\!\!\!\! d\mu\,\frac{\mu-\lambda}{\mu+\lambda} \left[(1+\sigma_j^{-2} \lambda^2)^{-\kappa+n-1}(1+\sigma_k^{-2} \mu^2)^{-\kappa+n-1}-(1+\sigma_j^{-2} \mu^2)^{-\kappa+n-1}(1+\sigma_k^{-2} \lambda^2)^{-\kappa+n-1}\right],
\end{align}
and 
\begin{align}
\nonumber
&\chi_{j,n+1}(r,s)=-\chi_{n+1,j}(r,s)=(1-\delta_{j,n+1})\int\limits_{(-\infty,r)\cup(s,\infty)} d\lambda \, (1+\sigma_j^{-2} \lambda^2)^{-\kappa+n-1} \\
&=\left[\frac{\sqrt{\pi } \,\sigma_j  \Gamma \left(\kappa-n +\frac{1}{2}\right)}{ \Gamma (\kappa-n +1)}+r \, _2F_1\left(\frac{1}{2},\kappa-n+1 ;\frac{3}{2};-\frac{r^2}{\sigma_j ^2}\right)-s \, _2F_1\left(\frac{1}{2},\kappa-n+1 ;\frac{3}{2};-\frac{s^2}{\sigma_j ^2}\right)\right](1-\delta_{j,n+1}).
\end{align}
Similarly, $\widetilde{\chi}_{j,k}(r,s)$ as given below yields the double gap probability using Eq.~\eqref{tErs2}:
\begin{align}
\nonumber
&\widetilde{\chi}_{j,k}(r,s)=\int_r^s d\lambda \int_r^s d\mu\, \frac{\mu-\lambda}{\mu+\lambda}\left[(1+\sigma_j^{-2} \lambda^2)^{-\kappa+n-1}(1+\sigma_k^{-2} \mu^2)^{-\kappa+n-1}\right] \\
&=\frac{1}{2}\int_r^s d\lambda \int_r^s d\mu \,\frac{\mu-\lambda}{\mu+\lambda} \left[(1+\sigma_j^{-2} \lambda^2)^{-\kappa+n-1}(1+\sigma_k^{-2} \mu^2)^{-\kappa+n-1}-(1+\sigma_j^{-2} \mu^2)^{-\kappa+n-1}(1+\sigma_k^{-2} \lambda^2)^{-\kappa+n-1}\right],
\end{align}
\begin{align}
\nonumber
&\widetilde{\chi}_{j,n+1}(r,s)=-\chi_{n+1,j}(r,s)=(1-\delta_{j,n+1})\int_r^s d\lambda \, (1+\sigma_j^{-2} \lambda^2)^{-\kappa+n-1} \\
&=\left[s \, _2F_1\left(\frac{1}{2},\kappa-n+1 ;\frac{3}{2};-\frac{s^2}{\sigma_j ^2}\right)-r \, _2F_1\left(\frac{1}{2},\kappa-n+1 ;\frac{3}{2};-\frac{r^2}{\sigma_j ^2}\right)\right](1-\delta_{j,n+1}).
\end{align}
The expressions for density of the extreme eigenvalues follow from Eqs.~\eqref{pS2} and~\eqref{pL2}. Similar to the correlated Gauss-Wigner case, here also we have $\chi(-\infty,-x)=\chi(x,\infty)$ which gives $E(-\infty,-x)=E(x,\infty)$ and $p_S(-x)=p_L(x)$. 

In Table~\ref{TabCL1Corr} we compare the gap probabilities obtained using analytical results and numerical simulations. Fig.~\ref{FigCL1Corr} shows the plots of distributions and densities for the extreme eigenvalues.


\subsection{Uncorrelated case}

The uncorrelated Cauchy-Lorentz (variant I) is obtained from Eq.~\eqref{CL1Corr} for $\bSg=\1$. We have for hermitian matrices $\bH$,
\begin{equation}
\label{CL1Un}
\mathcal{P}(\bH)\propto|\1+\bH^2|^{-\kappa},
\end{equation}
with $\kappa>n-1/2$. Interestingly, for $\kappa=n$ the matrices $\bH$ may be generated using $n$-dimensional unitary matrices $\bU$
\begin{equation}
\label{HU}
\bH=\frac{1}{\iota}\left(\frac{\1-\bU}{\1+\bU}\right),
\end{equation}
the measure $|\1+\bH^2|^{-n}d[\bH]$ being then equivalent to the Haar measure $d\mu(\bU)$~\cite{OSZ2010}. 
Here $\iota=\sqrt{-1}$ is the imaginary unit. For arbitrary $\kappa$, Eq.~\eqref{HU} can be used provided $\bU$ are taken from the measure $\boldsymbol{|}|\1+\bU| \boldsymbol{|}^{2(\kappa-n)}d\mu(\bU)$, where $\boldsymbol{|}|\cdot|\boldsymbol{|}$ stands for the absolute value of determinant.

Using Eq.~\eqref{CL1Un}, because of unitary-invariance, the joint probability density of eigenvalues can be immediately written down as
\begin{equation}
\label{CL1UnEv}
P(\{\lambda\})=C\prod_{l=1}^n\frac{1}{(1+\lambda_l^2)^\kappa}\cdot\Delta^2(\{\lambda\}).
\end{equation}
The normalization factor is provided by the result
\begin{align}
\nonumber
C^{-1}&=n!\,\Big|\frac{1+(-1)^{j+k}}{2}\,\Bt\Big(\frac{j+k-1}{2},\kappa-\frac{j+k-1}{2}\Big)\Big|_{j,k=1,...,n}\\
&=2^{n^2-2\kappa n+n}\pi^n\prod_{j=1}^n \frac{\Gamma(j+1)\Gamma(j+2\kappa-2n)}{\Gamma^2(j+\kappa-n)},
\end{align}
$\Bt(a,b)=\int_0^1 dt\,t^a(1-t)^b=\Gamma(a)\Gamma(b)/\Gamma(a+b)$ being the Beta function.

The uncorrelated Cauchy-Lorentz ensemble described by Eqs.~\eqref{CL1Un} or~\eqref{CL1UnEv} is known to exhibit Levy tails in its eigenvalue spectra. It has been investigated in the context of modeling financial correlations~\cite{BJNPZ2003,BJNPZ2004}. In the problem of quantum chaotic scattering, Cauchy-Lorentz density has played a crucial role in demonstrating equivalence between the Heidelberg (Hamiltonian) approach and the Mexico (Scattering matrix) approach, the latter being dictated by the Poisson kernel~\cite{Brouwer1995}. Gap probabilities for the uncorrelated case has been studied in~\cite{WF2000} and, for finite $n$ case, presented in terms of second-order second-degree ordinary differential equations which are related to certain Painlev\'{e}-VI transcendents. In~\cite{MSVV2012} the density of the largest eigenvalue has been calculated for large $n$ and its large deviations have been examined. We provide below explicit finite $n$ results as determinantal expressions.
\begin{table}
\caption{Gap probabilities: Comparison between analytical and simulation results for uncorrelated Cauchy-Lorentz (variant I) ensemble.}
\centering
\begin{tabular}{|c|c|c|c|c|c|c|c| }
\hline
\multirow{2}{*}{$n$} & \multirow{2}{*}{$\kappa$} & \multirow{2}{*}{$r$} & \multirow{2}{*}{$s$} & \multicolumn{2}{|c|}{$E(r,s)$}  & \multicolumn{2}{|c|}{$\widetilde{E}(r,s)$} \\ \cline{5-8}
& & & &Analytical  & Simulation &Analytical & Simulation \\
\hline\hline
2 & 3 & $-5$ & $-0.3$ & 0.2907 & 0.2904 & 0.0307 & 0.0309 \\
\hline
2 & 1.9 & $0.1$ & $4$ & 0.3159 & 0.3112 & 0.0555 & 0.0551 \\
\hline
3 & 3.5 & $-1$ & 0.6 & 0.0303 & 0.0300 & 0.0238 & 0.0239 \\ 
\hline
4 & 5.8 & $-0.65$  & $0.65$ & 0.0025 & 0.0028 & 0.0036 & 0.0038 \\ 
\hline
5 & 6 & $-7$ & 4.7 & 0.0000 & 0.0000 & 0.9194 & 0.9197\\
\hline
\end{tabular}
\label{TabCL1Un}
\end{table}
\begin{figure}[ht!]
\centering
 \includegraphics[width=0.75\textwidth]{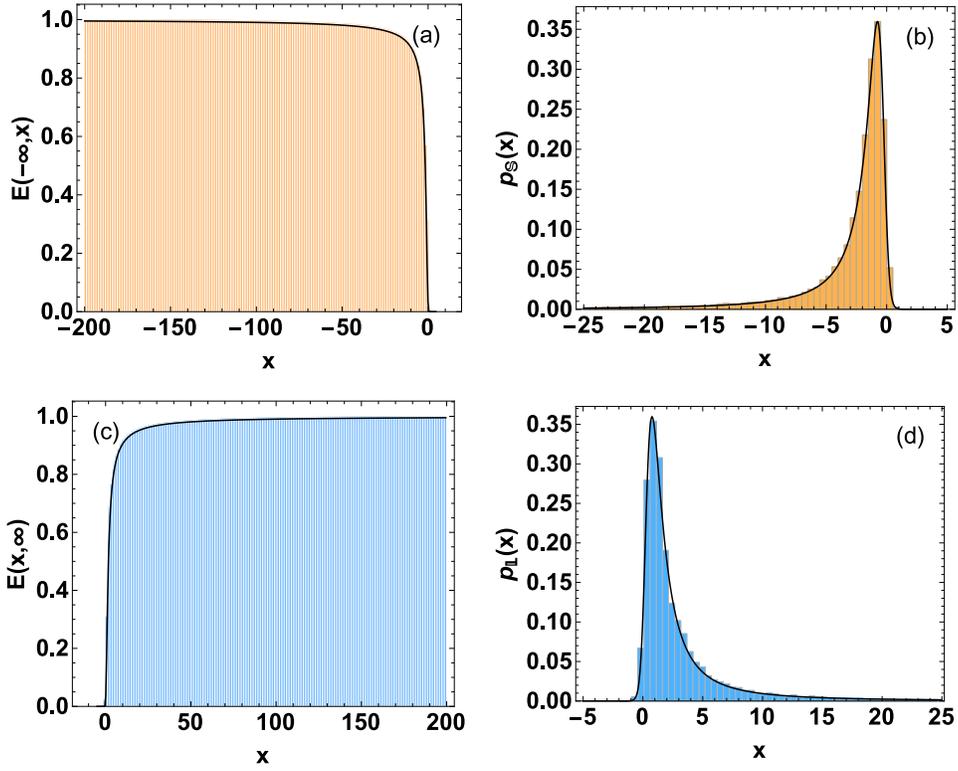}
\caption{Plots for uncorrelated Cauchy-Lorentz (variant I) ensemble with $n=3, \kappa=3$. (a) SF of the smallest eigenvalue (b) PDF of the smallest eigenvalue (c) CDF of the largest eigenvalue (d) PDF of the largest eigenvalue.}
\label{FigCL1Un}
\end{figure}

The expression for gap probability is given by Eq.~\eqref{Ers2} with 
\begin{equation}
\chi_{j,k}(r,s)
=\begin{cases}
\frac{(1+(-1)^{j+k})}{2}\,\Bt(\frac{j+k-1}{2},\kappa-\frac{j+k-1}{2})\\
+\frac{1}{2}(-1)^{j+k}\iota^{j+k-2\kappa-1}\left[\Bt(-\frac{1}{r^2},\kappa-\frac{j+k-1}{2},1-\kappa) -\Bt(-\frac{1}{s^2},\kappa-\frac{j+k-1}{2},1-\kappa)\right], & r<s<0,\\
\frac{1}{2}\iota^{j+k-2\kappa-1}\Big[(-1)^{j+k}\Bt(-\frac{1}{r^2},\kappa-\frac{j+k-1}{2},1-\kappa)+\Bt(-\frac{1}{s^2},\kappa-\frac{j+k-1}{2},1-\kappa)\Big], & r<0<s,\\
\frac{(1+(-1)^{j+k})}{2}\,\Bt(\frac{j+k-1}{2},\kappa-\frac{j+k-1}{2})\\
-\frac{1}{2}\iota^{j+k-2\kappa-1}\Big[\Bt(-\frac{1}{r^2},\kappa-\frac{j+k-1}{2},1-\kappa)  -\Bt(-\frac{1}{s^2},\kappa-\frac{j+k-1}{2},1-\kappa)\Big],& 0< r<s.
\end{cases}
\end{equation}
Here $\Bt(z,a,b)=\int_0^z dt\, t^{a-1}(1-t)^{b-1}$ is the incomplete Beta function.
We note that when $j+k$ is odd, all the three expressions above become identical. Similarly, for $\widetilde{E}(r,s)$ given by Eq.~\eqref{tErs2} we have
\begin{equation}
\widetilde{\chi}_{j,k}(r,s)
=\begin{cases}
\frac{1}{2}(-1)^{j+k}\iota^{j+k-2\kappa-1}\left[\Bt(-\frac{1}{s^2},\kappa-\frac{j+k-1}{2},1-\kappa) -\Bt(-\frac{1}{r^2},\kappa-\frac{j+k-1}{2},1-\kappa)\right], & r<s<0,\\
\frac{(1+(-1)^{j+k})}{2}\,\Bt(\frac{j+k-1}{2},\kappa-\frac{j+k-1}{2})\\
-\frac{1}{2}\iota^{j+k-2\kappa-1}\left[(-1)^{j+k}\Bt(-\frac{1}{r^2},\kappa-\frac{j+k-1}{2},1-\kappa) +\Bt(-\frac{1}{s^2},\kappa-\frac{j+k-1}{2},1-\kappa)\right], & r<0<s,\\
\frac{1}{2}\iota^{j+k-2\kappa-1}\left[\Bt(-\frac{1}{r^2},\kappa-\frac{j+k-1}{2},1-\kappa)-\Bt(-\frac{1}{s^2},\kappa-\frac{j+k-1}{2},1-\kappa)\right], & 0< r<s.
\end{cases}
\end{equation} 

For $r=-\infty$ or $s=\infty$ these reduce to
\begin{align}
\nonumber
&\chi_{j,k}(-\infty,x)=\widetilde{\chi}_{j,k}(x,\infty)\\
&=\begin{cases}
\frac{(1+(-1)^{j+k})}{2}\,\Bt(\frac{j+k-1}{2},\kappa-\frac{j+k-1}{2})-\frac{1}{2}(-1)^{j+k}\iota^{j+k-2\kappa-1}\Bt(-\frac{1}{x^2},\kappa-\frac{j+k-1}{2},1-\kappa), & x<0,\\
\frac{1}{2}\iota^{j+k-2\kappa-1}\Bt(-\frac{1}{x^2},\kappa-\frac{j+k-1}{2},1-\kappa), & x>0.
\end{cases}
\end{align} 
and
\begin{align}
\nonumber
&\chi_{j,k}(x,\infty)=\widetilde{\chi}_{j,k}(-\infty,x)\\
&=\begin{cases}
\frac{1}{2}(-1)^{j+k}\iota^{j+k-2\kappa-1}\Bt(-\frac{1}{x^2},\kappa-\frac{j+k-1}{2},1-\kappa), & x<0,\\
\frac{(1+(-1)^{j+k})}{2}\,\Bt(\frac{j+k-1}{2},\kappa-\frac{j+k-1}{2})-\frac{1}{2}\iota^{j+k-2\kappa-1}\Bt(-\frac{1}{x^2},\kappa-\frac{j+k-1}{2},1-\kappa) ,& x>0.
\end{cases}
\end{align} 
We remark that the kernels $\chi_{j,k}(r,s)$ and $\widetilde{\chi}_{j,k}(r,s)$ can also be represented in terms of Gauss hypergeometric function $\,_2F_1$, however we have opted to present the results in terms of Beta functions because of their simpler nature.

The expression for density of the smallest eigenvalue is obtained as Eq.~\eqref{pS1} with
\begin{equation}
\phi_{j,k}^{(i)}(x)
=\begin{cases}
x^{j+k-2}(1+x^2)^{-\kappa}, & j=i,\\
\chi_{j,k}(-\infty,x), & j\neq i.
 \end{cases}
 \end{equation}
Likewise the expressions for the density of the largest eigenvalue is given by Eq.~\eqref{pL1} with
\begin{equation}
\label{CL1Un_psi}
\psi_{j,k}^{(i)}(x)
=\begin{cases}
x^{j+k-2}(1+x^2)^{-\kappa}, & j=i,\\
\chi_{j,k}(x,\infty), & j\neq i.
 \end{cases}
 \end{equation}
Once again we have $E(-\infty,-x)=E(x,\infty)$ and $p_S(-x)=p_L(x)$, the reason being similar to that in the uncorrelated Gauss-Wigner case.

Table~\ref{TabCL1Un} collates the results for gap probabilities for various choice of parameters. Fig.~\ref{FigCL1Un} shows the comparison between the analytical and simulation-generated plots for extreme eigenvalue distribution and density functions.


\section{Cauchy-Lorentz ensemble: Variant II}
\label{SecCL2}

\subsection{Correlated case}
Consider $n\times n_A$ dimensional complex matrices $\bA$ from the density
\begin{equation}
\mathcal{P}_A(\bA)\propto|\1+\bSg^{-1}\bA\bA^\dag|^{-\kappa},
\end{equation}
with $n_A\geq n$, $\kappa> n_A+n-1$, and $\bSg=\diag(\sigma_1,...,\sigma_n), \sigma_j>0$. We are interested in the eigenvalues of the positive-definite Hermitian matrices $\bH=\bA\bA^\dag$, or equivalently the singular values of $\bA$. The probability density function satisfied by $\bH$ is
\begin{equation}
\label{CL2Corr0}
\mathcal{P}(\bH)\propto |\bH|^{n_A-n} |\1+\bSg^{-1}\bH|^{-\kappa}.
\end{equation}
We many generalize this density and replace $n_A-n$ by a real $\alpha>-1$, so that
\begin{equation}
\label{CL2Corr}
\mathcal{P}(\bH)\propto |\bH|^{\alpha} |\1+\bSg^{-1}\bH|^{-\kappa},
\end{equation}
with $\kappa>2n+\alpha-1$.
We refer to this as the correlated Cauchy-Lorentz (variant II) ensemble.

This ensemble admits construction via ratio of two  $n$-dimensional Laguerre-Wishart matrices, namely
\begin{equation}
\label{Ratio}
\bH=\frac{\bW_A}{\bW_B}.
\end{equation}
where $\bW_A$ and $\bW_B$ are respectively from the densities
\begin{equation}
\label{WisCorrAB}
P_A(\bW_A)\propto |\bW_A|^{\alpha}e^{-\tr \bSg_A^{-1}\bW_A},~~~~~
P_B(\bW_B)\propto |\bW_B|^{\beta}e^{-\tr \bSg_B^{-1}\bW_B},
\end{equation}
such that $\kappa=\alpha+\beta+2n$, and $\bSg^{-1}=\bSg_A^{-1}/\bSg_B^{-1}$. We have assumed here $\beta> -1$, similar to $\alpha$. A proof has been provided in~\ref{AppCL}. The above construction is particularly useful for generating matrices if $\alpha~(>-1)$ and $\kappa~(>2n+\alpha-1)$ happen to be integers. This is because in this case the $n$-dimensional Wishart matrices $\bW_A$ and $\bW_B$ with respective degrees of freedom $n_A,n_B$ can be easily generated with the aid of Eq.~\eqref{SGGdSd}, and then $\alpha=n_A-n, \kappa=n_A+n_B$.
We also note that the matrices $\bW_A \bW_B^{-1}$, $\bW_A^{-1} \bW_B$ and $\bW_B^{-1/2}\bW_A\bW_A^{-1/2}$ share the same set of eigenvalues as they correspond to the identical generalized eigenvalue problem. Therefore, as far as eigenvalue statistics is concerned, we may use the notation in Eq.~\eqref{Ratio} without any ambiguity. It is also clear from Eq.~\eqref{Ratio} that we are dealing with a multivariate generalization of beta distribution of the second kind.

The joint density of eigenvalues is obtained on employing the unitary group integral result given in~\cite{Orlov2004} as
\begin{equation}
\label{CL2CorrEv}
P(\{\lambda\})=C\,\Delta(\{\lambda\})\,\prod_{l=1}^n \lambda_j^{\alpha} \cdot|(1+\sigma_j^{-1} \lambda_k)^{-\kappa+n-1}|_{j,k=1,...,n},
\end{equation}
with $0\leq \lambda_j<\infty$.
Clearly we are dealing with type I ensemble as in Eq.~\eqref{Type1} with $w(\lambda)=\lambda^{\alpha}$, $f_j(\lambda_k)=\lambda_k^{j-1}$, and $g_j(\lambda_k)=(1+\sigma_j^{-1} \lambda_k)^{-\kappa+n-1}$.
The partition function is given by
\begin{align}
\label{CL2Corr_Norm}
\nonumber
C^{-1}&=n!\,\left|\sigma_k^{j+\alpha}\,\Bt(j+\alpha,\kappa-\alpha-n+1-j)\right|_{j,k=1,...,n} \\ 
&=n!\,\Delta(\{\sigma\})\prod_{j=1}^n \sigma_j^{\alpha+1}\,\Bt(j+\alpha,\kappa-\alpha-n+1-j).
\end{align}

The above variant of Cauchy-Lorentz model has been used in~\cite{WWKK2015} to work out the eigenvalue statistics of correlated Jacobi ensemble; see section~\ref{SecMANOVACorr}. Moreover, the JPDF in Eq.~\eqref{CL2CorrEv} has already been derived therein. We explore the behavior of its extreme eigenvalues below.

The expression for gap probability $E(r,s)$ is as in Eq.~\eqref{Ers1} with
\begin{align}
\chi_{j,k}(r,s)=\sigma_k^{j+\alpha}\left[(-1)^{-j-\alpha}\Bt\left(-\sigma_k^{-1}r,j+\alpha,n-\kappa\right)
-(-1)^{j+\alpha+n-\kappa}\,\Bt\left(-\sigma_k\, s^{-1},\kappa-\alpha-n+1-j,n-\kappa\right)\right].
\end{align}
While the double gap probability $\widetilde{E}(r,s)$ is obtained from Eq.~\eqref{tErs1} using
\begin{align}
\widetilde{\chi}_{j,k}(r,s)=(-1)^{-j-\alpha}\sigma_k^{j+\alpha}\left[\Bt\left(-\sigma_k^{-1}s,j+\alpha,n-\kappa\right)-\Bt\left(-\sigma_k^{-1}r,j+\alpha,n-\kappa\right)\right].
\end{align}
For $r=0$ or $s=\infty$ the above two equations simplify to
\begin{align}
\chi_{j,k}(0,x)=\widetilde{\chi}_{j,k}(x,\infty)=-(-1)^{j+\alpha+n-\kappa}\sigma_k^{j+\alpha}\,\Bt\left(-\sigma_k\, s^{-1},\kappa-\alpha-n+1-j,n-\kappa\right),
\end{align}
\begin{align}
\chi_{j,k}(x,\infty)=\widetilde{\chi}_{j,k}(0,x)=(-1)^{-j-\alpha}\sigma_k^{j+\alpha}\Bt\left(-\sigma_k^{-1}r,j+\alpha,n-\kappa\right).
\end{align}

\begin{table}
\caption{Gap probabilities: Comparison between analytical and simulation results for correlated Cauchy-Lorentz (variant II) ensemble. The $\sigma$ values are from $(\sigma_1,...,\sigma_5)=(1/2,2/3,7/6,3/10,3)$. The $n_A, n_B$ values are indicated whenever a straighforward matrix construction using Eq.~\eqref{Ratio} is possible.  }
\centering
\begin{tabular}{|c|c|c|c|c|c|c|c|c|c|c| }
\hline
\multirow{2}{*}{$n$} & \multirow{2}{*}{$\alpha$}  & \multirow{2}{*}{$\kappa$} & \multirow{2}{*}{$n_A$} & \multirow{2}{*}{$n_B$} & \multirow{2}{*}{$r$} & \multirow{2}{*}{$s$} & \multicolumn{2}{|c|}{$E(r,s)$}  & \multicolumn{2}{|c|}{$\widetilde{E}(r,s)$} \\ \cline{8-11}
& & & & &  & &Analytical  & Simulation &Analytical & Simulation \\
\hline\hline
2 & 1 & 5 & 3 & 2 & 0.4 & 6 & 0.2505 & 0.2517 & 0.1572 & 0.1563 \\
\hline
3 & $-0.6$ & 5.9 & $-$ & $-$ &0.23 & 5 & 0.0844 & 0.0822 & 0.0032 & 0.0032 \\ 
\hline
3 & 0 & 7 & 3 & 4 & 0.2 & 3 & 0.0501 & 0.0503 & 0.0130 &  0.0131 \\
\hline
4 & 1 & 9.5 & $-$ & $-$ & 0.2 & 5.3 & 0.0015 & 0.0014 & 0.0048 & 0.0043 \\
\hline
5 & 2 & 13 & 7 & 6& 3 & 5.5 & 0.5605 & 0.5608 & 0.0000 & 0.0000 \\
\hline
\end{tabular}
\label{TabCL2Corr}
\end{table}
\begin{figure}[t!]
\centering
 \includegraphics[width=0.75\textwidth]{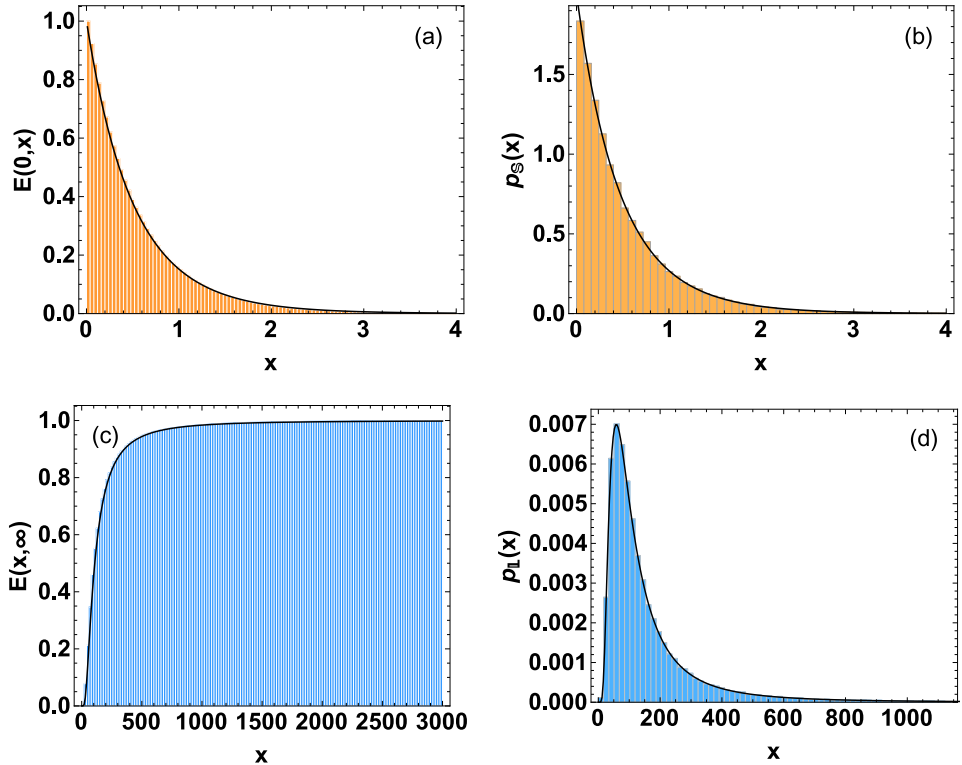}
\caption{Plots for correlated Cauchy-Lorentz (variant II) ensemble using $n=4, n_A=4, n_B= 5$ which gives $m=4, \alpha=9$. Also, $\bSg_A=\diag(33,21,15,43), \bSg_B=\diag(2,1,3,4)$ which gives $\bSg=\diag(33/2,21,5,43/4)$. (a) SF of the smallest eigenvalue (b) PDF of the smallest eigenvalue (c) CDF of the largest eigenvalue (d) PDF of the largest eigenvalue.}
\label{FigCL2Corr}
\end{figure}

The expression for density of the smallest eigenvalue is obtained as Eq.~\eqref{pS1} with
\begin{equation}
\phi_{j,k}^{(i)}(x)
=\begin{cases}
x^{j+\alpha-1}(1+\sigma_k^{-1}x)^{-\kappa+n-1}, & j=i,\\
\chi_{j,k}(0,x), & j\neq i.
 \end{cases}
 \end{equation}
Parallely, the expressions for density of the largest eigenvalue is given by Eq.~\eqref{pL1} with
\begin{equation}
\label{CL2Corr_psi}
\psi_{j,k}^{(i)}(x)
=\begin{cases}
x^{j+\alpha-1}(1+\sigma_k^{-1}x)^{-\kappa+n-1}, & j=i,\\
\chi_{j,k}(x,\infty), & j\neq i.
 \end{cases}
 \end{equation}

In Table~\ref{TabCL2Corr} we present the gap probability values for several combinations of parameters. Fig.~\ref{FigCL2Corr} exhibits the behavior of extreme eigenvalues.


\subsection{Uncorrelated case}

We now examine the uncorrelated version ($\bSg=\1$) of Eq.~\eqref{CL2Corr}, namely the density
\begin{equation}
\label{CL2Un}
\mathcal{P}(\bH)\propto |\bH|^{\alpha} |\1+\bH|^{-\kappa},
\end{equation}
involving positive-definite Hermitian matrices $\bH$ with $\kappa>2n+\alpha-1$. In accordance with the correlated case, this ensemble is related to
\begin{equation}
\label{CL2UnAAd}
\mathcal{P}_A(\bA)\propto|\1+\bA\bA^\dag|^{-\kappa},
\end{equation}
with $\bA$ being a rectangular complex matrix. The form~\eqref{CL2UnAAd} of uncorrelated Cauchy-Lorentz ensemble has been explored in connection with the projection formula in supersymmetry~\cite{KKG2014}.

Similar to the correlated case, this ensemble can be realized using the ratio of two Wishart matrices, i.e., $\bH=\bW_A/\bW_B$
where now $\bW_A$ and $\bW_B$ are from the distributions
\begin{equation}
\label{WisUnAB}
P_A(\bW_A)\propto |\bW_A|^{\alpha}e^{-\tr \bW_A},~~~~~
P_B(\bW_B)\propto |\bW_B|^{\beta}e^{-\tr \bW_B},
\end{equation}
such that $\kappa=\alpha+\beta+2n$. As in the correlated case, for integer $\alpha~(>-1)$ and $\kappa~(>2n+\alpha-1)$ the above construction is particularly useful. The matrix model $\bH=\bW_A/\bW_B$ follows in a special limit ($a=b\rightarrow \infty$) of a more general case $a\bW_A/(\1+b \bW_B); a,b>0$ considered in~\cite{Kumar2015}.
\begin{table}
\caption{Gap probabilities: Comparison between analytical and simulation results for uncorrelated Cauchy-Lorentz (variant II) ensemble. The $n_A, n_B$ values are indicated whenever an easy matrix construction is possible using ratio of two Wishart matrices.}
\centering
\begin{tabular}{|c|c|c|c|c|c|c|c|c|c|c| }
\hline
\multirow{2}{*}{$n$} & \multirow{2}{*}{$\alpha$}  & \multirow{2}{*}{$\kappa$} & \multirow{2}{*}{$n_A$} & \multirow{2}{*}{$n_B$} & \multirow{2}{*}{$r$} & \multirow{2}{*}{$s$} & \multicolumn{2}{|c|}{$E(r,s)$}  & \multicolumn{2}{|c|}{$\widetilde{E}(r,s)$} \\ \cline{8-11}
& & & & & & &Analytical  & Simulation &Analytical & Simulation \\
\hline\hline
2 & 0 & 5 & 2 & 3 & 0.9 & 5.6 & 0.3169 & 0.3134 & 0.0122 & 0.0120 \\
\hline
3 & $-0.8$ & 6.7 & $-$ & $-$ & 0.3 & 11.3 & 0.0157 & 0.0159 & 0.0011 & 0.0010 \\
\hline
3 & 1 & 9 &4 &5 & 0.6 & 24 & 0.0028 & 0.0025 & 0.0096 & 0.0092 \\
\hline 
4 & 1 & 11 & 5 & 6 & 3.4 & $\infty$ & 0.2118 & 0.2111 & 0.0000 & 0.0000 \\ 
\hline
5 & 2 & 13.7 & $-$ &$-$& 0&3.5&0.0000 & 0.0000 & 0.0155 & 0.0155\\
\hline
\end{tabular}
\label{TabCL2Un}
\end{table}
\begin{figure}[t!]
\centering
 \includegraphics[width=0.75\textwidth]{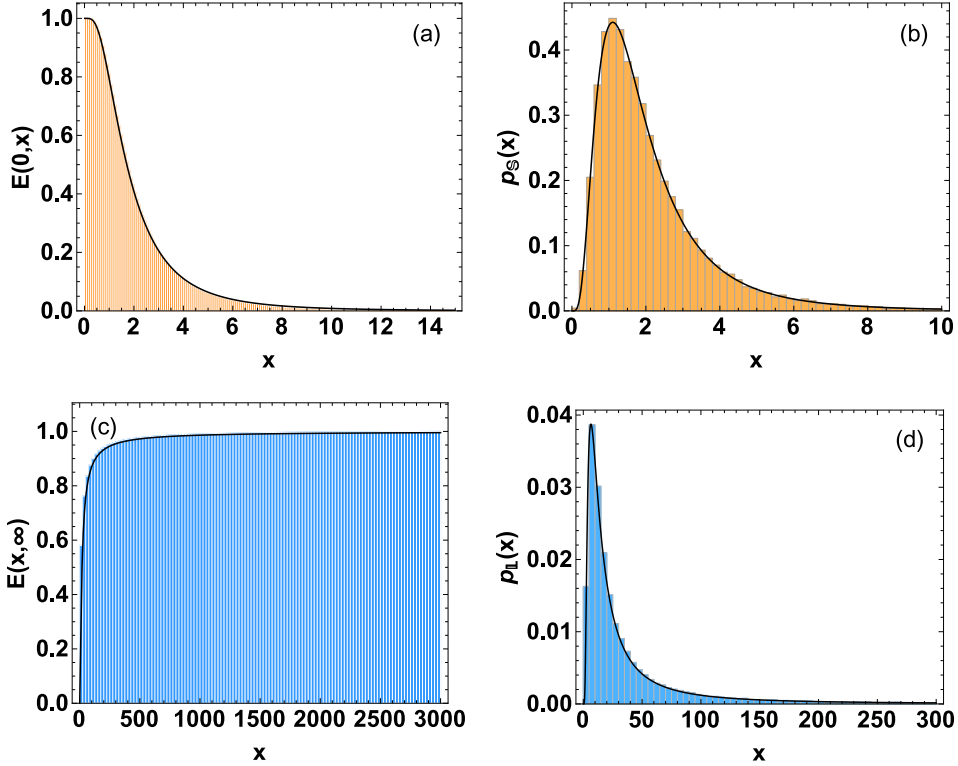}
\caption{Plots for uncorrelated Cauchy-Lorentz (variant II) ensemble using $n=2, n_A=7, n_B= 2$ which gives $\alpha=5, \kappa=9$. (a) SF of the smallest eigenvalue (b) PDF of the smallest eigenvalue (c) CDF of the largest eigenvalue (d) PDF of the largest eigenvalue.}
\label{FigCL2Un}
\end{figure}

The joint density of eigenvalues follows readily because of unitarily-invariant nature of Eq.~\eqref{CL2Un}. We have
\begin{equation}
P(\{\lambda\})=C\,\Delta^2(\{\lambda\})\prod_{j=1}^n \lambda_j^{\alpha}(1+\lambda_j)^{-\kappa}.
\end{equation}
The normalization factor is obtained using
\begin{align}
\nonumber
C^{-1}&=n!\,|\Bt(j+k+\alpha-1,\kappa-\alpha-j-k+1)|_{j,k=1,...,n}\\ 
&=n!\,\prod_{j=1}^n \frac{\Gamma(j)\Gamma(j+\alpha)\Gamma(\kappa-\alpha-n-j+1)}{\Gamma(\kappa-j+1)}.
\end{align}
The gap probabilities are given by Eqs.~\eqref{Ers1} and~\eqref{tErs1} with
\begin{equation}
\chi_{j,k}(r,s)=-(-1)^{-j-k-\alpha}\,\Bt\left(-r,\alpha+j+k-1,1-\kappa\right)
-(-1)^{j+k+\alpha-\kappa}\,\Bt\left(-s^{-1},\kappa-\alpha-j-k+1,1-\kappa\right),
\end{equation}
and,
\begin{align}
\widetilde{\chi}_{j,k}(r,s)=(-1)^{-j-k-\alpha}\,\Big[\Bt\left(-r,\alpha+j+k-1,1-\kappa\right)-\Bt(-s,\alpha+j+k-1,1-\kappa)\Big].
\end{align}
These simplify to the following for $r=0$ or $s=\infty$:
\begin{equation}
\chi_{j,k}(0,x)=\widetilde{\chi}_{j,k}(x,\infty)=-(-1)^{j+k+\alpha-\kappa}\,\Bt\left(-x^{-1},\kappa-\alpha-j-k+1,1-\kappa\right),
\end{equation}
\begin{equation}
\chi_{j,k}(x,\infty)=\widetilde{\chi}_{j,k}(0,x)=-(-1)^{-j-k-\alpha}\,\Bt\left(-x,\alpha+j+k-1,1-\kappa\right).
\end{equation}

The expression for density of the smallest eigenvalue is obtained as Eq.~\eqref{pS1} with
\begin{equation}
\phi_{j,k}^{(i)}(x)
=\begin{cases}
x^{\alpha+j+k-2}\,(1+x)^{-\kappa}, & j=i,\\
\chi_{j,k}(0,x), & j\neq i.
 \end{cases}
 \end{equation}
while, the result for density of the largest eigenvalue is given by Eq.~\eqref{pL1} with
\begin{equation}
\psi_{j,k}^{(i)}(x)
=\begin{cases}
x^{\alpha+j+k-2}\,(1+x)^{-\kappa}, & j=i,\\
\chi_{j,k}(x,\infty), & j\neq i.
 \end{cases}
 \end{equation}

 In Table~\ref{TabCL2Un} we collect the gap probability values for various choice of parameters. Fig.~\ref{FigCL2Un} shows the plots of distributions and densities obtained from analytical expressions (solid lines) as well as from Monte-Carlo simulations (histograms).


\section{Jacobi-MANOVA ensemble}
\label{SecMANOVA}

\subsection{Correlated case}
\label{SecMANOVACorr}

Consider the matrix probability density
\begin{equation}
\label{MANOVACorr}
\mathcal{P}(\bH)\propto |\bH|^{\alpha}\,|\1-\bH|^{\beta}\,|\1+(\bSg^{-1}-\1)\bH |^{-\kappa},
\end{equation}
with $\bH$ being $n\times n$-dimensional matrices, such that $\mathbb{0}\leq \bH \leq \1$, and $\bSg=\diag(\sigma_1,...,\sigma_n), \sigma_j>0$. By the notation $\bH\ge \bK$ for matrices, we mean that $\bH-\bK$ is non-negative definite.
Also $\alpha,\beta>-1$ and $\kappa$ is any real parameter. Eq.~\eqref{MANOVACorr} then defines the correlated Jacobi-MANOVA ensemble of random matrices. The term MANOVA derives from the area of multivariate statistics and stands for Multivariate Analysis Of VAriance, while Jacobi has to do with the occurrence of Jacobi weight function in the JPDF of eigenvalues; see below. We note that the eigenvalues for matrix model~\eqref{MANOVACorr} varies from 0 to 1. In contrast, if we consider the density
\begin{equation}
\label{JacobiCorr}
\mathcal{P}(\bH)\propto |\1+\bH|^{\alpha}\,|\1-\bH|^{\beta}\,|(\bSg^{-1}+\1)+(\bSg^{-1}-\1)\bH |^{-\kappa},
\end{equation}
then the corresponding eigenvalues range from $-1$ to 1. The two models,~\eqref{MANOVACorr} and ~\eqref{JacobiCorr}, are trivially related by a shift and a scaling transformations, and so are the corresponding eigenvalues.

For $\beta=\kappa-\alpha-2n>-1$ the correlated Jacobi-MANOVA ensemble has a direct relationship with the Cauchy-Lorentz (variant II) ensemble, Eq.~\eqref{CL2Corr}. For if we consider matrices $\widetilde{\bH}$ drawn from Eq.~\eqref{CL2Corr}, then the matrices
\begin{equation}
\label{MANOVA-CL2}
\bH=\frac{\widetilde{\bH}}{\1+\widetilde{\bH}}
\end{equation}
are distributed according to the density~\eqref{MANOVACorr}. Equivalently, if we consider matrices the matrices $\bH$ drawn from Eq.~\eqref{MANOVACorr}, then the matrices
\begin{equation}
\label{MANOVA-CL}
\widetilde{\bH}=\frac{\bH}{\1-\bH},
\end{equation}
are distributed according to the density~\eqref{CL2Corr}. As a consequence of this correspondence, we may construct the Jacobi-MANOVA matrices as follows. Consider $n$-dimensional Laguerre-Wishart matrices $\bW_A$ and $\bW_B$ drawn from the densities as in Eq.~\eqref{WisCorrAB}, then
\begin{equation}
\label{DoubleWishart}
\bH= \frac{\bW_A}{\bW_A+\bW_B}
\end{equation}
constitutes the ensemble described by Eq.~\eqref{MANOVACorr}, with $\alpha,\beta$ as in Eq.~\eqref{WisCorrAB} and $\kappa=\alpha+\beta+2n$. Also, $\bSg^{-1}=\bSg_A^{-1}/\bSg_B^{-1}$. In~\ref{AppMANOVACorr} we provide a proof. Again, for (permitted) integer values of $\alpha,\beta,\kappa$ the above construction gives a very accessible way of generating Jacobi-MANOVA matrices, as the $n$-dimensional Wishart matrices $\bW_A,\bW_B$ with respective degrees of freedom $n_A,n_B$ can be easily constructed using Eq.~\eqref{SGGdSd}. In this case the parameters are given by $\alpha=n_A-n,\beta=n_B-n$ and $\kappa=n_A+n_B$. From Eq.~\eqref{DoubleWishart} it is also clear that we are dealing with a multivariate generalization of beta distribution of the first kind. Correlated Jacobi-MANOVA ensemble has been introduced very recently and solved exactly to obtain the joint eigenvalue density and correlation function of arbitrary order~\cite{WWKK2015}.

A unitary group integral result from~\cite{Orlov2004} leads to the joint probability density of eigenvalues for matrix density~\eqref{MANOVACorr} as
\begin{align}
P(\lambda_1,...,\lambda_n)=C \Delta(\{\lambda\})\,\prod_{l=1}^n \lambda_l^{\alpha}(1-\lambda_l)^{\beta}\Big| [1+(\sigma_j^{-1}-1)\lambda_k]^{-\kappa+n-1}\Big|_{j,k=1,...,n}.
\end{align}
Note that for $\kappa-n+1=0$, one encounters $0/0$ form in the JPDF of eigenvalues as well as in the expressions for gap probabilities and extreme eigenvalue densities, therefore a limiting  procedure must be invoked. However, this case (along with $\bSg=\1$) corresponds to the uncorrelated Jacobi-MANOVA ensemble for which much simpler results exist, as given in the next section. Another case when a limiting procedure has to be followed occurs when $\kappa-n+1=-1$ with $n>2$.

As a consequence of the relationship with Cauchy Lorentz (variant II), when $\beta=\kappa-\alpha-2n$ the results for correlated Jacobi-MANOVA can be written down easily by a simple variable transformation. This also includes the case when construction of Jacobi-MANOVA matrices is possible using two Laguerre-Wishart matrices, as described above. Therefore, this scenario is of special interest and we explicitly deal with this case along with the general $\kappa$ case.


 \subsubsection{Arbitrary $\kappa$, and $\alpha,\beta>-1$}
 The partition function is obtained as
 \begin{equation}
 \label{MANOVACorr_Norm}
 C^{-1}=n!\,\big|\Bt(j+\alpha,\beta+1)\,_2F_1(j+\alpha,\kappa-n+1;j+\alpha+\beta+1;1-\sigma_k^{-1})\big|_{j,k=1,...,n},
 \end{equation}
 where $_2F_1(a,b;c;z)$ is the Gauss hypergeometric function, as already mentioned.
 
 The gap probability is given by Eq.~\eqref{Ers1} with
 \begin{align}
 \nonumber
 \chi_{j,k}(r,s)&=\Bt(j+\alpha,\beta+1)\,_2F_1(j+\alpha,\kappa-n+1;j+\alpha+\beta+1;1-\sigma_k^{-1})\\
 \nonumber
 &+\frac{r^{j+\alpha}}{j+\alpha}\, F_1\left(j+\alpha ;-\beta ,\kappa-n+1;j+\alpha +1;r,(1-\sigma _k^{-1})r\right)\\
& -\frac{s^{j+\alpha}}{j+\alpha}\, F_1\left(j+\alpha ;-\beta ,\kappa-n+1;j+\alpha +1;s,(1-\sigma _k^{-1})s\right),
 \end{align}
where $F_1(a;b_1,b_2;c;x,y)=\frac{\Gamma(c)}{\Gamma(a)\Gamma(b_1)\Gamma(b_2)}\sum_{m,n=0}^\infty\frac{\Gamma(a+m+n)\Gamma(b_1+m)\Gamma(b_2+n)\,x^m y^n}{m!\, n!\,\Gamma(c+m+n)}$ represents the Appell hypergeometric function of two variables.
Similarly, we have
 \begin{align}
 \nonumber
 \widetilde{\chi}_{j,k}(r,s)& =\frac{s^{j+\alpha}}{j+\alpha}\, F_1\left(j+\alpha ;-\beta ,\kappa-n+1;j+\alpha +1;s,(1-\sigma _k^{-1})s\right)\\
&-\frac{r^{j+\alpha}}{j+\alpha}\, F_1\left(j+\alpha ;-\beta ,\kappa-n+1;j+\alpha +1;r,(1-\sigma _k^{-1})r\right).
 \end{align}
For $r=0$ or $s=1$, we have the following simplifications:
 \begin{align}
 \nonumber
 \chi_{j,k}(0,x)=\widetilde{\chi}_{j,k}(x,1)&=\Bt(j+\alpha,\beta+1)\,_2F_1\left(j+\alpha,\kappa-n+1;j+\alpha+\beta+1;1-\sigma_k^{-1}\right)\\
& -\frac{x^{j+\alpha}}{j+\alpha}\, F_1\left(j+\alpha ;-\beta ,\kappa-n+1;j+\alpha +1;x,(1-\sigma _k^{-1})x\right),
 \end{align}
  \begin{align}
 \nonumber
 \chi_{j,k}(x,1)=\widetilde{\chi}_{j,k}(0,x)&=\frac{x^{j+\alpha}}{j+\alpha}\, F_1\left(j+\alpha ;-\beta ,\kappa-n+1;j+\alpha +1;x,(1-\sigma _k^{-1})x\right).\\
 \end{align}
 
 The density of smallest eigenvalue is given by Eq.~\eqref{pS1} with
\begin{equation}
\phi_{j,k}^{(i)}(x)
=\begin{cases}
x^{j+\alpha-1}(1-x)^\beta(1+(\sigma_k^{-1}-1)x)^{-\kappa+n-1}, & j=i,\\
 \chi_{j,k}(0,x), & j\neq i,
 \end{cases}
 \end{equation}
and that of the largest eigenvalue by Eq.~\eqref{pL1} with
\begin{equation}
\psi_{j,k}^{(i)}(x)
=\begin{cases}
x^{j+\alpha-1}(1-x)^\beta(1+(\sigma_k^{-1}-1)x)^{-\kappa+n-1}, & j=i,\\
 \chi_{j,k}(x,1), & j\neq i.
 \end{cases}
 \end{equation}


 \subsubsection{Arbitrary $\alpha>-1$, and $\beta=\kappa-\alpha-2n>-1$}
 
In view of the result~\eqref{MANOVA-CL2} we consider the transformation~\cite{WWKK2015}
\begin{equation}
\lambda_k=\frac{\mu_k}{1+\mu_k},~~~~k=1,...,n.
\end{equation}
The joint density in the new variables is
\begin{align}
\nonumber
\widetilde{P}(\mu_1,...,\mu_n)=\widetilde{C}  \Delta(\{\mu\})\, \prod_{l=1}^n\mu_l^{\alpha}\cdot\Big| (1+\sigma_j^{-1}\mu_k)^{-\kappa+n-1}\Big|_{j,k=1,...,n}.
\end{align}
Here the normalization factor $\widetilde{C}$ is same as the $C$ in Eq.~\eqref{CL2Corr_Norm}. Also, $C$ of Eq.~\eqref{MANOVACorr_Norm} reduces to $\widetilde{C}$ when $\beta=\kappa-\alpha-2n$.
We use Eqs.~\eqref{CL2Corr_Norm}$-$\eqref{CL2Corr_psi} to write the expressions for gap probabilities and extreme eigenvalue densities in the $\mu$ variables, and then transform back to the $\lambda$ variables. We wind up with the results as given below.

\begin{table}
\caption{Gap probabilities: Comparison between analytical and simulation results for correlated Jacobi-MANOVA ensemble. The $\sigma$ values are from $(\sigma_1,...,\sigma_5)=(3/2,5/6,7/4,2/5,4)$. The $n_A, n_B$ values are indicated whenever a straightforward matrix construction using Eq.~\eqref{DoubleWishart} follows. }
\centering
\begin{tabular}{|c|c|c|c|c|c|c|c|c|c|c|c| }
\hline
\multirow{2}{*}{$n$} & \multirow{2}{*}{$\alpha$}& \multirow{2}{*}{$\beta$}& \multirow{2}{*}{$\kappa$} & \multirow{2}{*}{$n_A$}& \multirow{2}{*}{$n_B$}& \multirow{2}{*}{$r$} & \multirow{2}{*}{$s$} & \multicolumn{2}{|c|}{$E(r,s)$}  & \multicolumn{2}{|c|}{$\widetilde{E}(r,s)$} \\ \cline{9-12}
& & & & & & & &Analytical  & Simulation &Analytical & Simulation \\
\hline\hline
2 & 1 & 2 & 7 & 3& 4 & 0.2 & 0.6 & 0.3128 & 0.3133 & 0.1071 & 0.1065  \\
\hline
2 & $-0.1$ & $-0.3$ & 2 & $-$ & $-$ & 0 & 0.5 & 0.0891 & 0.0883 & 0.0352 &  0.0356 \\ 
\hline
3 & 1.3 & 1.4 & 1.5 & $-$ & $-$ & 0.5 & 1 & 0.0029 & 0.0026 & 0.0034 & 0.0035 \\
\hline
4 & 1 & 2 & 11 & 5 & 6 & 0.1 & 0.7 & 0.0013 & 0.0013 & 0.0058 & 0.0054  \\
\hline
5 & 2.2 & 4.5 & 0 & $-$ & $-$ & 0.15 & 0.9 & 0.0000 & 0.0000 & 0.0702 & 0.0704  \\
\hline
\end{tabular}
\label{TabMANOVACorr}
\end{table}
\begin{figure}[t!]
\centering
 \includegraphics[width=0.62\textwidth]{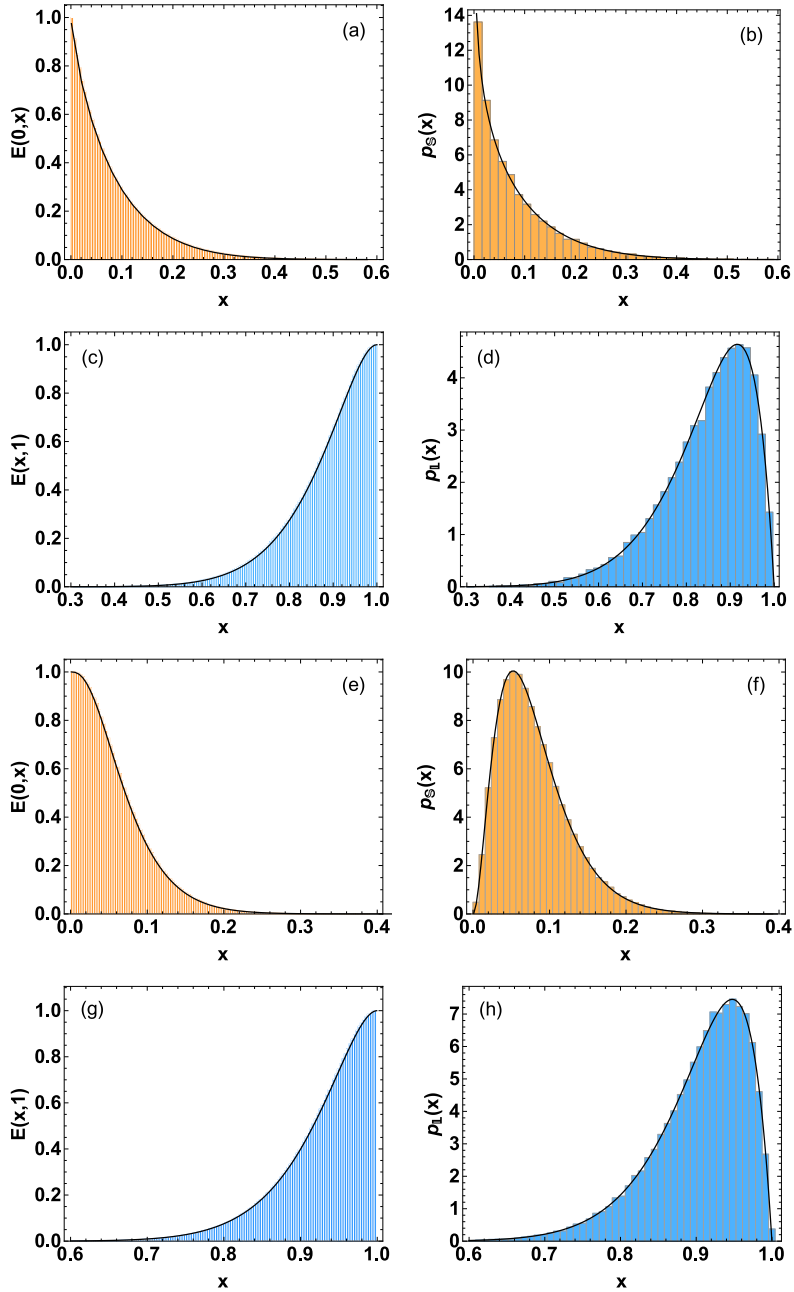}
\caption{Plots for correlated Jacobi-MANOVA ensemble. (a),(e) SF of the smallest eigenvalue; (b),(f) PDF of the smallest eigenvalue; (c),(g) CDF of the largest eigenvalue; (d),(h) PDF of the largest eigenvalue. For (a)-(d), the parameter values used are $n=3,\alpha=-0.2,\beta=1,\kappa=2.5$. Also, $\bSg=(4,7/5,10/3)$. This set of values does not admit a straightforward matrix construction using Eq.~\eqref{DoubleWishart}, and therefore results from section 7.1.1 have been used for the analytical plots. For (e)-(h) the parameter set is such that matrix construction using Eq.~\eqref{DoubleWishart} is easily possible and analytical results from section 7.1.2 have been used. The parameter values are  $n=4,n_A=6, n_B=5$, and $\bSg_A=(5/2,1/11,2/7,1/4)$, $\bSg_B=(7/6,1/5,4/3,1/3)$. These lead to $\alpha=2,\beta=1,\kappa=11$, and $\bSg=(15/7,5/11,3/14,3/4)$. }
\label{FigMANOVACorr}
\end{figure}

The gap probability is given by Eq.~\eqref{Ers1} with
\begin{align}
\chi_{j,k}(r,s)=\sigma_k^{j+\alpha}\left[(-1)^{-j-\alpha}\Bt\left(\frac{r}{\sigma_k(r-1)},j+\alpha,n-\kappa\right)-(-1)^{j+\alpha+n-\kappa}\,\Bt\left(\frac{\sigma_k\,(s-1)}{s},\kappa-\alpha-n+1-j,n-\kappa\right)\right].
\end{align}
While, the double gap probability is obtained using Eq.~\eqref{tErs1} with
\begin{align}
\widetilde{\chi}_{j,k}(r,s)=(-1)^{-j-\alpha}\sigma_k^{j+\alpha}\left[\Bt\left(\frac{s}{\sigma_k(s-1)},j+\alpha,n-\kappa\right)-\Bt\left(\frac{r}{\sigma_k(r-1)},j+\alpha,n-\kappa\right)\right].
\end{align}
For $r=0$ or $s=\infty$ these reduce to
\begin{align}
\chi_{j,k}(0,x)=\widetilde{\chi}_{j,k}(x,\infty)=-(-1)^{j+\alpha+n-\kappa}\sigma_k^{j+\alpha}\,\Bt\left(\frac{\sigma_k\,(s-1)}{s},\kappa-\alpha-n+1-j,n-\kappa\right),
\end{align}
\begin{align}
\chi_{j,k}(x,1)=\widetilde{\chi}_{j,k}(0,x)=(-1)^{-j-\alpha}\sigma_k^{j+\alpha}\,\Bt\left(\frac{r}{\sigma_k(r-1)},j+\alpha,n-\kappa\right).
\end{align}

The probability density function for the smallest eigenvalue is obtained from Eq.~\eqref{pS1} after incorporating Jacobian of transformation as
 \begin{align}
p_{\mathbb{S}}(x)&=n!\,\widetilde{C}\,\dfrac{1}{(1-x)^2}\sum_{i=1}^n |\phi_{j,k}^{(i)}(x)|_{j,k=1,...,n},
\end{align}
with
\begin{equation}
\phi_{j,k}^{(i)}(x)
=\begin{cases}
\left(\dfrac{x}{1-x}\right)^{j+\alpha-1}\left(1+\dfrac{x}{\sigma_k(1-x)}\right)^{-\kappa+n-1}, & j=i,\\
\chi_{j,k}(0,x), & j\neq i.
 \end{cases}
 \end{equation}
The expressions for density of the largest eigenvalue is given by 
\begin{align}
p_{\mathbb{L}}(x)&=n!\,\widetilde{C}\,\frac{1}{(1-x)^2}\sum_{i=1}^n |\psi_{j,k}^{(i)}(x)|_{j,k=1,...,n},
\end{align}
with
\begin{equation}
\psi_{j,k}^{(i)}(x)
=\begin{cases}
\left(\dfrac{x}{1-x}\right)^{j+\alpha-1}\left(1+\dfrac{x}{\sigma_k(1-x)}\right)^{-\kappa+n-1}, & j=i,\\
\chi_{j,k}(x,1), & j\neq i.
 \end{cases}
 \end{equation}

In Table~\ref{TabMANOVACorr} we compare the gap probability values obtained from analytical formulae and from numerical simulations. Fig.~\ref{FigMANOVACorr} shows the plots of distribution functions and probability density functions. For (a)-(d), we have used the parameter values $n=3,\alpha=-0.2,\beta=1,\kappa=2.5$. Also, the correlation matrix taken is $\bSg=(4,7/5,10/3)$. This set of values does not admit matrix construction using Eq.~\eqref{DoubleWishart} in a straightforward manner, and therefore we use the results from section 7.1.1 for the analytical plots. For (e)-(h) the parameter set is such that matrix construction using Eq.~\eqref{DoubleWishart} is trivially possible and we use the analytical results from section 7.1.2. The parameter values are $n=4,n_A=6, n_B=5$, and $\bSg_A=(5/2,1/11,2/7,1/4)$, $\bSg_B=(7/6,1/5,4/3,1/3)$. These lead to $\alpha=2,\beta=1,\kappa=11$, and $\bSg=(15/7,5/11,3/14,3/4)$. We find perfect agreement between the analytic predictions and the simulation results.  

We note that $E(0,x)$, $E(x,1)$, $p_\mathbb{S}(x)$ and $p_\mathbb{L}(x)$ for a given set of  $n,\alpha, \beta,\bSg$ are, respectively, same as $E(1-x,1)$, $E(0,1-x)$, $p_\mathbb{L}(1-x)$ and $p_\mathbb{S}(1-x)$ for $n,\beta,\alpha,\bSg^{-1}$. This correspondence has to do with the structure of the matrix density~\eqref{MANOVACorr}.


\subsection{Uncorrelated case}

The uncorrelated Jacobi-MANOVA case follows from Eq.~\eqref{MANOVACorr} for $\bSg=\1$. We have
\begin{equation}
\label{MANOVAUn}
\mathcal{P}(\bH)\propto|\bH|^{\alpha}\,|\1-\bH|^{\beta}\,
\end{equation}
where $\alpha,\beta>-1$. The eigenvalues of this model are from  $0$ to $1$.
On the other hand if we work with the model
\begin{equation}
\label{JacobiUn}
\mathcal{P}(\bH)\propto|\1+\bH|^{\alpha}\,|\1-\bH|^{\beta},
\end{equation}
then the eigenvalues lie in $[-1,1]$. Again, the two matrix models are related by a simple linear transformation and so are the eigenvalues. 
As in the correlated case, the matrices from~\eqref{MANOVAUn} can be generated using $n$-dimensional Laguerre-Wishart matrices as $\bH= \bW_A/(\bW_A+\bW_B)$
where $\bW_A,\bW_B$ are from~\eqref{WisUnAB}. 

The joint eigenvalue density is obtained as
\begin{equation}
\label{MANOVAUnEv}
P(\lambda_1,...,\lambda_N)=C\Delta^2(\{\lambda\}) \prod_{j=1}^n \lambda_j^{\alpha}(1-\lambda_j)^{\beta}. 
\end{equation}
This JPDF is trivially related to the standard Jacobi ensemble of random matrices~\cite{Mehta2004,Forrester2010} via the linear transformation $\lambda_j\rightarrow (1+\lambda_j)/2$.
The normalization factor is obtained using
\begin{align}
\nonumber
C^{-1}&=n!\,|\Bt(j+k+\alpha-1,\beta+1)|_{j,k=1,...,n}\\
&=\prod_{j=1}^n\frac{\Gamma(j+1)\,\Gamma(j+\alpha)\,\Gamma(j+\beta)}{\Gamma(j+\alpha+\beta+n)}.
\end{align}
Jacobi-MANOVA ensemble has an important role in multivariate statistics~\cite{Anderson2003,Muirhead2005}. Moreover, it find applications in quantum conductance problem~\cite{Forrester2006,SS2006,SSW2008,KSS2009,KP2010b,KP2010c,VMB2010,MS2013,SMM2014,CFV2015} and multiple channel optical fiber communication~\cite{DFS2013,KMV2014}. The extreme eigenvalue statistics for uncorrelated Jacobi-MANOVA has been investigated for large $n$ in~\cite{Johnstone2008,RKC2012}. Finite $n$ case has been considered in~\cite{KD2008}. However, even for the complex case the results have been provided in terms of hypergeometric function of matrix argument. We provide below expressions which are in terms of standard Beta functions, and are much easier to implement. 

The expression for gap probability for the eigenvalues described by Eq.~\eqref{MANOVAUnEv} is given by Eq.~\eqref{Ers2} with 
\begin{equation}
\chi_{j,k}(r,s)=\Bt(j+k+\alpha-1,\beta+1)+\Bt(r;j+k+\alpha-1,\beta+1)-\Bt(s;j+k+\alpha-1,\beta+1).
\end{equation} 
Similarly,
\begin{equation}
\widetilde{\chi}_{j,k}(r,s)=\Bt(s;j+k+\alpha-1,\beta+1)-\Bt(r;j+k+\alpha-1,\beta+1)
\end{equation} 
leads to the expression for double gap probability using Eq.~\eqref{tErs2}. These simplify to the following for $r=0$ or $s=1$:
\begin{equation}
\chi_{j,k}(0,x)=\widetilde{\chi}_{j,k}(x,1)=\Bt(j+k+\alpha-1,\beta+1)-\Bt(x;j+k+\alpha-1,\beta+1),
\end{equation}
\begin{equation}
\chi_{j,k}(x,1)=\widetilde{\chi}_{j,k}(0,x)=\Bt(x;j+k+\alpha-1,\beta+1).
\end{equation}
\begin{table}
\caption{Gap probabilities: Comparison between analytical and simulation results for uncorrelated Jacobi-MANOVA. The $n_A,n_B$ values are indicated whenever an easy matrix construction is possible using $\bH=\bW_A/(\bW_A+\bW_B)$. }
\centering
\begin{tabular}{|c|c|c|c|c|c|c|c|c|c|c|}
\hline
\multirow{2}{*}{$n$} & \multirow{2}{*}{$\alpha$}& \multirow{2}{*}{$\beta$}& \multirow{2}{*}{$n_A$}& \multirow{2}{*}{$n_B$} & \multirow{2}{*}{$r$} & \multirow{2}{*}{$s$} & \multicolumn{2}{|c|}{$E(r,s)$}  & \multicolumn{2}{|c|}{$\widetilde{E}(r,s)$} \\ \cline{8-11}
& & & & & & &Analytical  & Simulation &Analytical & Simulation \\
\hline\hline
2 & 3 & 1 & 5 & 3 & 0.3 & 0.7 & 0.1977 & 0.1970 & 0.0850 & 0.0855 \\
\hline
3 & $-0.4$ & 3.5 & $-$ & $-$ & 0.2 & 0.8 & 0.0179 & 0.0178 & 0.0029 & 0.0028 \\ 
\hline
3 & 0 & 4 & 3 & 7 & 0.05 & 0.4 & 0.0555 & 0.0538 & 0.0146 & 0.0141 \\
\hline
4 & 3.2 & 1.8 & $-$ & $-$ & 0 & 0.37 & 0.0233 & 0.0232 & 0.0000 &  0.0000 \\
\hline
5 & 2 & 9 & 14 & 7 & 0.5 & 1 & 0.0000 & 0.0000 & 0.0135 & 0.0130 \\
\hline
\end{tabular}
\label{TabMANOVAUn}
\end{table}
\begin{figure}[t!]
\centering
 \includegraphics[width=0.75\textwidth]{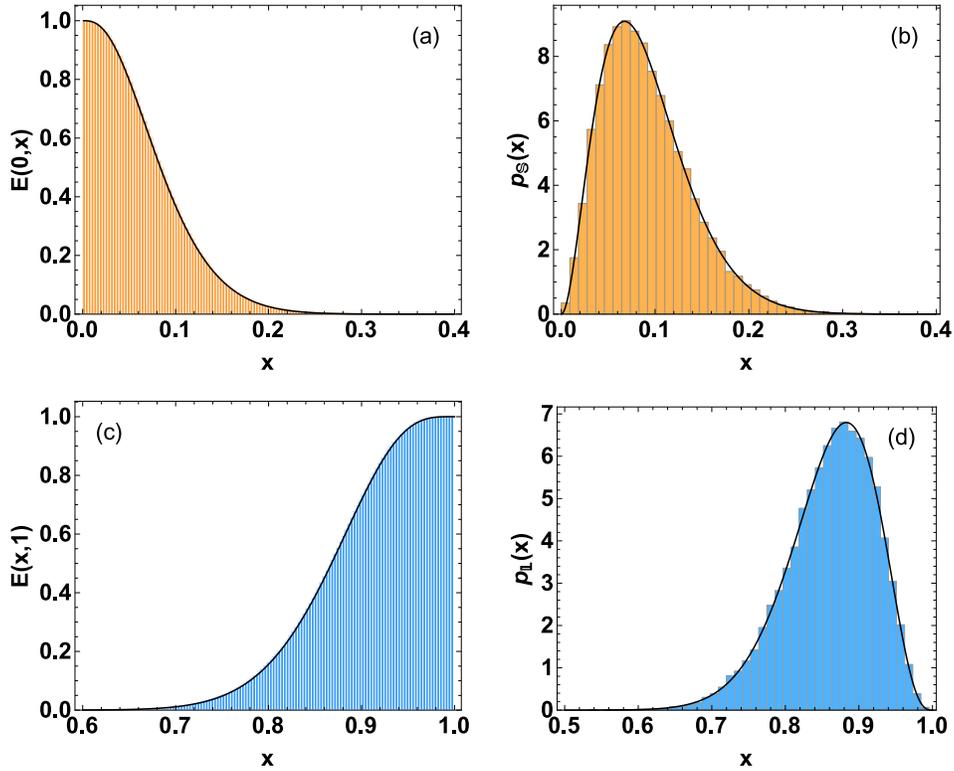}
\caption{Plots for uncorrelated Jacobi-MANOVA ensemble with $n=5,n_A=7, n_B=8$. (a) SF of the smallest eigenvalue, (b) PDF of the smallest eigenvalue, (d) CDF of the largest eigenvalue, (d) PDF of the largest eigenvalue.}
\label{FigMANOVAUn}
\end{figure}

The probability density function for the smallest eigenvalue is given by Eq.~\eqref{pS1} with
\begin{equation}
\phi_{j,k}^{(i)}(x)
=\begin{cases}
x^{\alpha+j+k-2}(1-x)^{\beta}, & j=i,\\
\chi_{j,k}(0,x), & j\neq i.
 \end{cases}
 \end{equation}
Likewise, the expression for the probability density of the largest eigenvalue is obtained as Eq.~\eqref{pL1} with
 \begin{align}
\label{pswig}
p_{\mathbb{L}}(x)&=n!\,C\,\sum_{i=1}^n |\psi_{j,k}^{(i)}(x)|_{j,k=1,...,n},
\end{align}
where
\begin{equation}
\psi_{j,k}^{(i)}(x)
=\begin{cases}
x^{\alpha+j+k-2}(1-x)^{\beta}, & j=i,\\
\chi_{j,k}(x,1), & j\neq i.
 \end{cases}
 \end{equation}

In Table~\ref{TabMANOVAUn} we compare the results for gap probabilities for various choice of parameters. Fig.~\ref{FigMANOVAUn} shows the plots of extreme eigenvalue distributions and densities obtained from analytical expressions as well as from Monte-Carlo simulation.
 
Similar to the correlated case, $E(0,x)$, $E(x,1)$, $p_\mathbb{S}(x)$ and $p_\mathbb{L}(x)$ for a given set of  $n,\alpha, \beta$ are, respectively, same as $E(1-x,1)$, $E(0,1-x)$, $p_\mathbb{L}(1-x)$ and $p_\mathbb{S}(1-x)$ for $n,\beta,\alpha$. 

\section{Bures-Hall ensemble}
\label{SecBH}

\subsection{Correlated case}
Bures-Hall measure usually pertains to a fixed trace scenario, its origin being in the field of quantum information~\cite{OSZ2010,Hall1998,Slater1999,SZ2003,SZ2004}. However, we retain the name in the case of unrestricted trace also. We can write down the matrix probability density for positive-definite Hermitian matrices $\bH$ as
\begin{equation}
\label{BHCorr}
\mathcal{P}(\bH)\propto |\bH|^{\alpha}e^{-\tr\bSg^{-1}\bH}\frac{1}{\Delta_{+}(\bH)},
\end{equation}
where $\alpha>-1$ is a real parameter,  $\bSg=\diag(\sigma_1,...,\sigma_n), \sigma_j>0$, and $\Delta_{+}(\bH)\equiv\Delta_{+}(\{\lambda\})$, $\lambda$'s being the eigenvalues. We can realize the above ensemble using
\begin{equation}
\label{BHWisCL1}
\bH=\frac{\bW}{\1+\bSg^{1/2}\bG^2\bSg^{1/2}},
\end{equation}
where the $\bW$'s are $n$-dimensional matrices from the complex Wishart distribution $\mathcal{P}(\bW)\propto |\bW|^\eta \exp(-\tr \bSg^{-1}\bW)$, as in Eq.~\eqref{WisCorr}, but with $\eta>-1/2$. While the $\bG$'s are $n$-dimensional Hermitian matrices from correlated Cauchy-Lorentz distribution (Variant I),
\begin{equation}
\label{CL1}
\mathcal{P}(\bG)\propto|\1+\bSg\bG^2|^{-(\eta+n)},
\end{equation}
cf. Eq.~\eqref{CL1Corr}.
The parameter $\alpha$ of Eq.~\eqref{BHCorr} is then given by $\alpha=\eta-1/2$. A proof has been outlined in~\ref{AppBH}. A special case occurs when the $\bW$ are $n$-dimensional Wishart matrices with degree of freedom $m$, so that $\eta=m-n$. In this case $\alpha~(>-1)$ assumes half-integer values, namely $\alpha=-1/2,1/2,3/2$ etc.

The joint density of eigenvalues turns out to be
\begin{equation}
P(\{\lambda\})=C\frac{\Delta(\{\lambda\})}{\Delta_+(\{\lambda\})}\prod_{l=1}^n \lambda_l^\alpha\,\cdot |e^{-\sigma_j^{-1}\lambda_k}|_{j,k=1,...,n},
\end{equation}
with $\lambda_j\geq0$. This JPDF is of type II since it can be rewritten as the product of a Pfaffian and a determinant, as in the correlated Gauss-Wigner case.
The partition function is obtained as
\begin{equation}
\nonumber
C^{-1}=n!\pf[h_{jk}]_{j,k=1,..,N},
\end{equation}
where $N$ is as defined in Eq.~\eqref{DefN}. For even $n$ we have
\begin{equation}
h_{j,k}=\frac{\Gamma^2(\alpha+1)}{2}\left[\sigma_j^{2\alpha+2}\,_2F_1(2\alpha+2,\alpha+2;2\alpha+3;1-\sigma_k^{-1}\sigma_j)-\sigma_k^{2\alpha+2}\,_2F_1(2\alpha+2,\alpha+2;2\alpha+3;1-\sigma_j^{-1}\sigma_k)\right].
\end{equation}
and when $n$ is odd, in addition we have
\begin{equation}
h_{j,n+1}=-h_{n+1,j}=\sigma_j^{\alpha+1}\Gamma(\alpha+1)(1-\delta_{j,n+1}).
\end{equation}
For $\alpha=-1/2$, $C^{-1}$ simplifies to
\begin{align}
C^{-1}&=\begin{cases}
n!\pf\left[\pi\sqrt{\sigma_k}\sqrt{\sigma_j}(\sqrt{\sigma_k}-\sqrt{\sigma_j})/(\sqrt{\sigma_k}+\sqrt{\sigma_j})\right]_{j,k=1,...,n}, & n \text{ even},\\
n!\pf\begin{bmatrix}\left[\pi\sqrt{\sigma_k}\sqrt{\sigma_j}(\sqrt{\sigma_k}-\sqrt{\sigma_j})/(\sqrt{\sigma_k}+\sqrt{\sigma_j})\right]_{j,k=1,...,n} & \left[\sqrt{\pi}\sqrt{\sigma_j}\right]_{j=1,...,n} \\
\left[-\sqrt{\pi}\sqrt{\sigma_k}\right]_{k=1,...,n} & 0 \end{bmatrix}, & n \text{ odd}.
\end{cases}\\
&=n!\,\pi^{n/2}\prod_{i=1}^n \sqrt{\sigma_i}\,\cdot \prod_{j>k}\frac{\sqrt{\sigma_j}-\sqrt{\sigma_k}}{\sqrt{\sigma_j}+\sqrt{\sigma_k}}.
\end{align}

\begin{table}
\caption{Gap probabilities: Comparison between analytical and simulation results for correlated Bures-Hall ensemble. The $\sigma$ values are from $(\sigma_1,...,\sigma_5)=(3/7,8/9,7/11,2/5,3)$. The $m$ values are indicated when $\bW$ in Eq.~\eqref{BHWisCL1} are Wishart matrices with degree of freedom $m$. }
\centering
\begin{tabular}{|c|c|c|c|c|c|c|c|c| }
\hline
\multirow{2}{*}{$n$} & \multirow{2}{*}{$\alpha$} & \multirow{2}{*}{$m$} & \multirow{2}{*}{$r$} & \multirow{2}{*}{$s$} & \multicolumn{2}{|c|}{$E(r,s)$}  & \multicolumn{2}{|c|}{$\widetilde{E}(r,s)$} \\ \cline{6-9}
& & & & &Analytical  & Simulation &Analytical & Simulation \\
\hline\hline
2 & $-0.5$ & 2 &0.5 & 5 & 0.2263 & 0.2268 & 0.0229 & 0.0216 \\
\hline
2 & $3$ &$-$ &2.9 & 7 & 0.2603 & 0.2605 & 0.0153 & 0.0158 \\
\hline
3 & 3.5 & 7 & 0 & 3.2 & 0.0011 & 0.0011 & 0.0134 & 0.0138 \\ 
\hline
4 & 0 & $-$ & 1.3 & 5.2 & 0.0511 & 0.0530 & 0.0000 & 0.000 \\
\hline
5 & 2.3 & $-$ & 0.6 & 11.7 & 0.0000 & 0.0000 & 0.0361 & 0.0368 \\
\hline
\end{tabular}
\label{TabBuresCorr}
\end{table}

\begin{figure}[ht!]
\centering
 \includegraphics[width=0.75\textwidth]{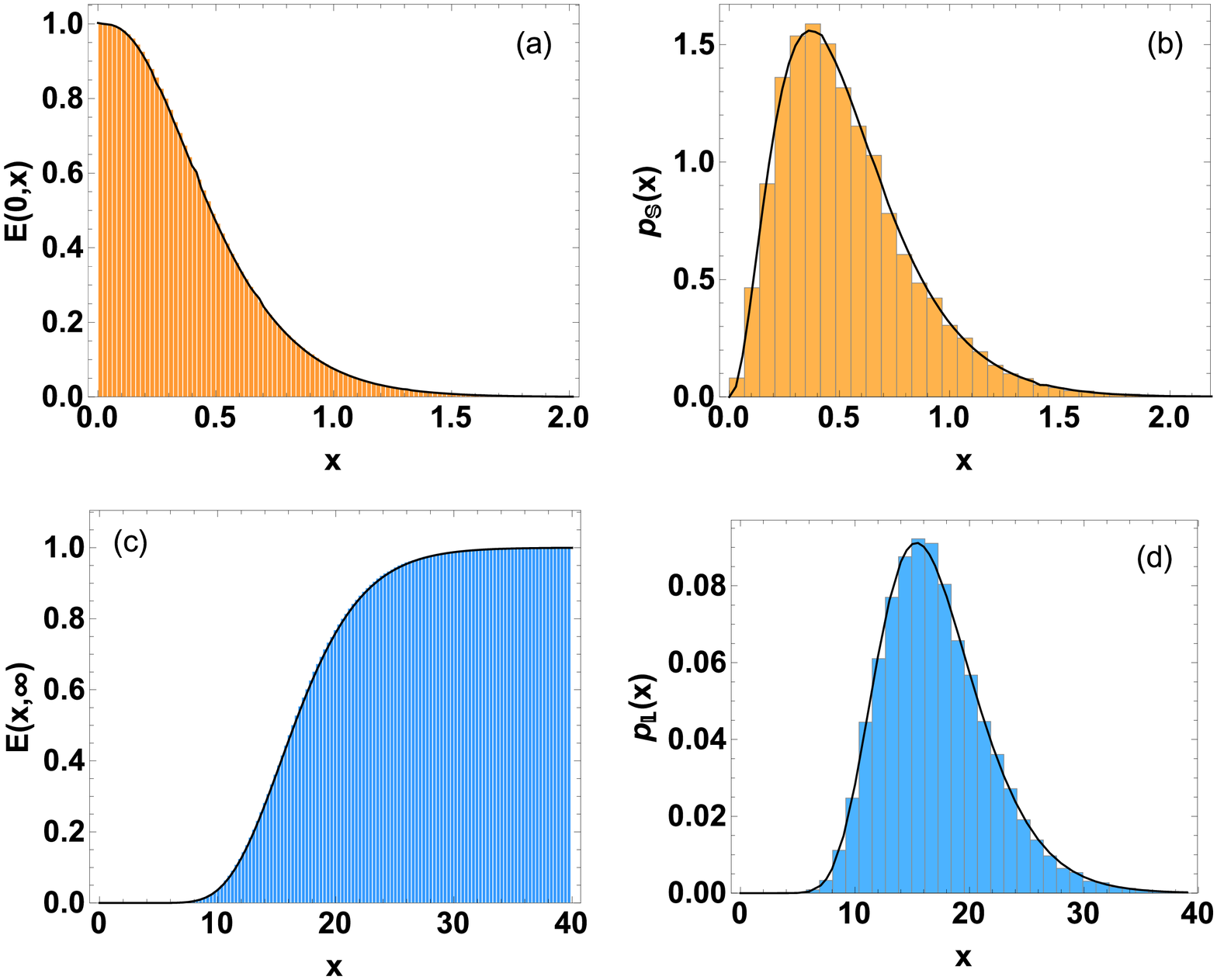}
\caption{Plots for correlated Bures-Hall ensemble with $n=5$, $\alpha=16/7$ and $\bSg=(4/11,3/4,4/3,8/5,9/4)$. (a) SF of the smallest eigenvalue, (b) PDF of the smallest eigenvalue, (d) CDF of the largest eigenvalue, (d) PDF of the largest eigenvalue.}
\label{FigBuresCorr}
\end{figure}

Gap probability $E(r,s)$ is given by Eqs.~\eqref{Ers2} with
\begin{align}
\nonumber
\chi_{j,k}(r,s)&=\int\limits_{(0,r)\cup(s,\infty)} d\lambda \int\limits_{(0,r)\cup(s,\infty)} d\mu\, \lambda^\alpha \mu^\alpha \,\frac{\mu-\lambda}{\mu+\lambda}\,e^{-\sigma_j^{-1}\lambda}e^{-\sigma_k^{-1}\mu} \\
&=\frac{1}{2}\int\limits_{(0,r)\cup(s,\infty)} d\lambda \int\limits_{(0,r)\cup(s,\infty)} d\mu \, \lambda^\alpha\,\mu^\alpha\,\frac{\mu-\lambda}{\mu+\lambda} \left(e^{-\sigma_j^{-1}\lambda}e^{-\sigma_k^{-1}\mu}-e^{-\sigma_j^{-1}\mu}e^{-\sigma_k^{-1}\lambda}\right),
\end{align}
and
\begin{align}
\chi_{j,n+1}(r,s)=-\chi_{n+1,j}(r,s)&=(1-\delta_{j,n+1})\int\limits_{(0,r)\cup(s,\infty)} d\lambda \, \lambda^\alpha e^{-\sigma_j^{-1}\lambda} \\
&=\sigma_j^{\alpha+1}\left[\gamma(\alpha+1,\sigma_j^{-1}r)+\Gamma(\alpha+1,\sigma_j^{-1}s)\right](1-\delta_{j,n+1}).
\end{align}

The double-gap probability $\widetilde{E}(r,s)$ is given by Eq.~\eqref{tErs2} with
\begin{align}
\nonumber
\widetilde{\chi}_{j,k}(r,s)&=\int_r^s d\lambda \int_r^s d\mu\, \lambda^\alpha \mu^\alpha\, \frac{\mu-\lambda}{\mu+\lambda}\,  e^{-\sigma_j^{-1}\lambda}e^{-\sigma_k^{-1}\mu}\\
&=\frac{1}{2}\int_r^s d\lambda \int_r^s d\mu \, \lambda^\alpha \mu^\alpha \,\frac{\mu-\lambda}{\mu+\lambda}\, \left(e^{-\sigma_j^{-1}\lambda}e^{-\sigma_k^{-1}\mu}-e^{-\sigma_j^{-1}\mu}e^{-\sigma_k^{-1}\lambda}\right),
\end{align}
and
\begin{align}
\nonumber
\widetilde{\chi}_{j,n+1}(r,s)=-\chi_{n+1,j}(r,s)&=(1-\delta_{j,n+1})\int_r^s d\lambda\, \lambda^{\alpha} \, e^{-\sigma_j^{-1}\lambda} \\
&=\sigma_j^{\alpha+1}\left[\Gamma(\alpha+1,\sigma_j^{-1}r)-\Gamma(\alpha+1,\sigma_j^{-1}s)\right](1-\delta_{j,n+1}).
\end{align}

 In Table~\ref{TabBuresCorr} we list the gap probability values obtained from analytical results and from numerical simulation. Fig.~\ref{FigBuresCorr} shows behavior of extreme eigenvalues in terms of distributions and densities.


\subsection{Uncorrelated case}

The matrix probability density in this case is
\begin{equation}
\label{BHUn}
\mathcal{P}(\bH)\propto |\bH|^{\alpha}e^{-\bH}\frac{1}{\Delta_{+}(\bH)},
\end{equation}
for positive-definite Hermitian matrices $\bH$, with $\alpha>-1$. Similar to the correlated case we can associate to Eq.~\eqref{BHUn} the matrix model
\begin{equation}
\label{BHUnWisCL1}
\bH=\frac{\bW}{\1+\bG^2},
\end{equation}
where $\bW$ is an $n$-dimensional uncorrelated Laguerre-Wishart matrix as in Eq.~\eqref{WisUn} with the replacement $\alpha\rightarrow \eta~(>-1/2)$, and $\bG$ is $n$-dimensional Hermitian matrix from uncorrelated Cauchy-Lorentz (variant I) density, i.e., from Eq.~\eqref{CL1} with $\bSg=\1$. The parameter $\alpha$ of Eq.~\eqref{BHUn} is then given by $\eta-1/2$. Again, if the $\bW$'s are $n$-dimensional Wishart matrices with degree of freedom $m$, then $\alpha$ assumes the half-integer values decided by $m-n-1/2$.
In this case, using the results in~\cite{OSZ2010}, we know that $\bH$ can also be generated using $\bW$'s and $n$-dimensional unitary matrices $\bU$ from the measure $\boldsymbol{|}|\1+\bU| \boldsymbol{|}^{2(m-n)}d\mu(\bU)$ as
\begin{equation}
\bH=\frac{(\1+\bU)}{2}\bW\frac{(\1+\bU^\dag)}{2}.
\end{equation}
For $m=n$, i.e. $\alpha=-1/2$, the measure $\boldsymbol{|}|\1+\bU| \boldsymbol{|}^{2(m-n)}d\mu(\bU)$ simplifies to $d\mu(\bU)$, which is the Haar measure on the group of $n$-dimensional unitary matrices.

The joint density of eigenvalues is obtained as
\begin{equation}
P(\{\lambda\})=C\frac{\Delta(\{\lambda\})}{\Delta_+(\{\lambda\})}\prod_{l=1}^n \lambda^\alpha e^{-\lambda}\,\Delta(\{\lambda\}),
\end{equation}
where the partition function is given by
\begin{align}
C^{-1}&=\begin{cases}
n!\,\pf\left[\dfrac{k-j}{j+k+2\alpha}\Gamma(j+\alpha)\Gamma(k+\alpha)\right]_{j,k=1,...,n}, & n \text{ even},\\
n!\,\pf\begin{bmatrix}\left[\dfrac{k-j}{j+k+2\alpha}\Gamma(j+\alpha)\Gamma(k+\alpha)\right]_{j,k=1,...,n} & \left[\Gamma(j+\alpha)\right]_{j=1,...,n} \\
\left[-\Gamma(k+\alpha)\right]_{k=1,...,n} & 0 \end{bmatrix}, & n \text{ odd}.
\end{cases}\\
&=\frac{\pi^{n/2}}{2^{n^2+2\alpha n}}\prod_{i=1}^n \frac{\Gamma(j+1)\,\Gamma(j+2\alpha+1)}{\Gamma(j+\alpha+1/2)}.
\end{align}
The above ensemble has been recently shown to be related to the Cauchy two-matrix ensemble in~\cite{FK2015}.
It is worth a mention here that the normalization factor for fixed trace Bures-Hall ensemble had remained elusive for long~\cite{Hall1998,Slater1999}, and was finally worked out in~\cite{SZ2003}. Since the normalization factor for fixed trace uncorrelated Bures-Hall ensemble has a simple relationship (via Laplace transform) with the normalization factor for unrestricted trace uncorrelated Bures-Hall ensemble, the above result readily leads to the former.

\begin{table}
\caption{Gap probabilities: Comparison between analytical and simulation results for uncorrelated Bures-Hall ensemble. The $m$ values are indicated when $\bW$ in Eq.~\eqref{BHUnWisCL1} are Wishart matrices with degree of freedom $m$.}
\centering
\begin{tabular}{|c|c|c|c|c|c|c|c|c|c| }
\hline
\multirow{2}{*}{$n$} & \multirow{2}{*}{$\alpha$}& \multirow{2}{*}{$m$} & \multirow{2}{*}{$r$} & \multirow{2}{*}{$s$} & \multicolumn{2}{|c|}{$E(r,s)$}  & \multicolumn{2}{|c|}{$\widetilde{E}(r,s)$} \\ \cline{6-9}
& & & &  &Analytical  & Simulation &Analytical & Simulation \\
\hline\hline
2 & $-0.5$ & 2 & 0.3 & 4 & 0.0925 & 0.0920 & 0.1314 & 0.1311 \\
\hline
3 & 0 & $-$ & 1 & 10.5 & 0.0049 & 0.0048 & 0.0098 & 0.0093 \\ 
\hline
3 & 1.7 & $-$ & 2.3 & 8.3 & 0.0586 & 0.0574 & 0.0078 & 0.0072 \\ 
\hline
4 & 1.5 & 6 & 1.4 & 8.3 & 0.0014 & 0.0013 & 0.0046 & 0.0047 \\
\hline
5 & 3.5 & 9 & 8 & 12 & 0.3019 & 0.3015 & 0.0000 & 0.0000 \\
\hline
\end{tabular}
\label{TabBuresUn}
\end{table}
\begin{figure}[ht!]
\centering
 \includegraphics[width=0.75\textwidth]{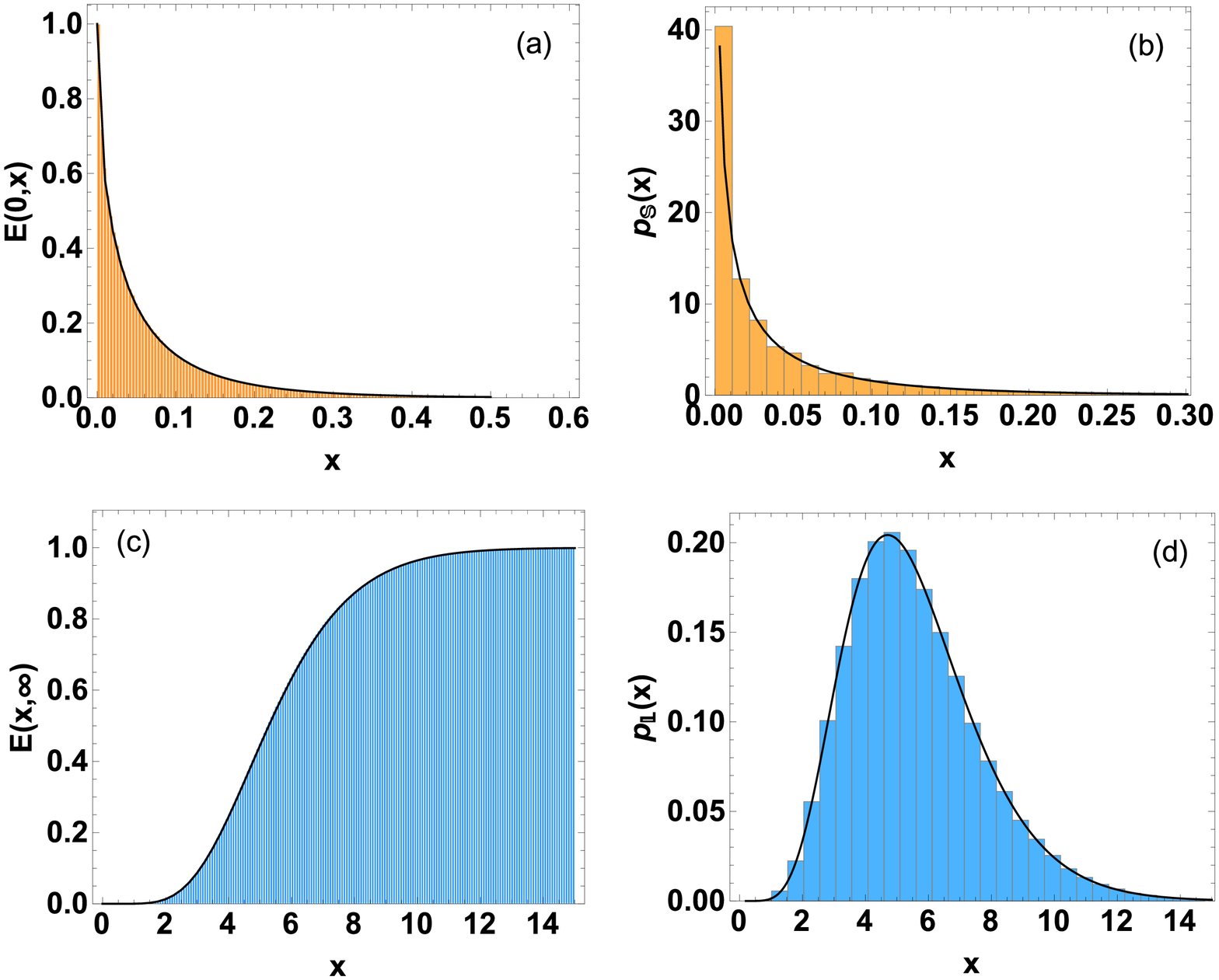}
\caption{Plots for uncorrelated Bures-Hall ensemble with $n=4$ and $\alpha=-1/2$. (a) SF of the smallest eigenvalue, (b) PDF of the smallest eigenvalue, (d) CDF of the largest eigenvalue, (d) PDF of the largest eigenvalue.}
\label{FigBuresUn}
\end{figure}

The gap probability is given by Eq.~\eqref{Ers2} with
\begin{align}
\nonumber
\chi_{j,k}(r,s)&=\int\limits_{(0,r)\cup(s,\infty)} d\lambda \int\limits_{(0,r)\cup(s,\infty)} d\mu\, \frac{\mu-\lambda}{\mu+\lambda}\,e^{-\lambda}e^{-\mu}\lambda^{j+\alpha-1} \mu^{k+\alpha-1} \\
&=\frac{1}{2}\int\limits_{(0,r)\cup(s,\infty)} d\lambda \int\limits_{(0,r)\cup(s,\infty)} d\mu \,\frac{\mu-\lambda}{\mu+\lambda}\, e^{-\lambda}e^{-\mu}\lambda^{\alpha-1}\,\mu^{\alpha-1} \left(\lambda^j\mu^k-\mu^j\lambda^k\right),
\end{align}
and
\begin{align}
\nonumber
\chi_{j,n+1}(r,s)=-\chi_{n+1,j}(r,s)&=(1-\delta_{j,n+1})\int\limits_{(0,r)\cup(s,\infty)} d\lambda \, \lambda^{\alpha+j-1} e^{-\lambda} \\
&=\left[\gamma(j+\alpha,r)+\Gamma(j+\alpha,s)\right](1-\delta_{j,n+1}).
\end{align}
The double-gap probability is given by Eq.~\eqref{tErs2} with
\begin{align}
\nonumber
\widetilde{\chi}_{j,k}(r,s)&=\int_r^s d\lambda \int_r^s d\mu\, \frac{\mu-\lambda}{\mu+\lambda}\,e^{-\lambda}e^{-\mu}\lambda^{j+\alpha-1} \mu^{k+\alpha-1}\\
&=\frac{1}{2}\int_r^s d\lambda \int_r^s d\mu \,\frac{\mu-\lambda}{\mu+\lambda}\,e^{-\lambda}e^{-\mu} \lambda^{\alpha-1}\,\mu^{\alpha-1} \left(\lambda^j\mu^k-\mu^j\lambda^k\right),
\end{align}
and
\begin{align}
\nonumber
\widetilde{\chi}_{j,n+1}(r,s)=-\widetilde{\chi}_{n+1,j}(r,s)&=(1-\delta_{j,n+1})\int_r^s d\lambda\, \lambda^{\alpha+j-1} e^{-\lambda}  \\
&=\left[\Gamma(j+\alpha,r)-\Gamma(j+\alpha,s)\right](1-\delta_{j,n+1}).
\end{align}

In Table~\ref{TabBuresUn} we compare the results for gap probabilities for several parameter choices. Fig.~\ref{FigBuresUn} shows the plots for distribution functions and the densities for the smallest eigenvalue as well as the largest eigenvalue.

\section{Summary and Outlook}
\label{SecConc}

In this work we derived exact results for gap probabilities and probability densities of the extreme eigenvalues of six matrix ensembles which are of central importance to random matrix theory. We considered both correlated and uncorrelated variants of these ensembles. In the process we also proposed generalized versions of some of these ensembles and came across some interesting relationships. All the analytical results for gap probabilities and probability densities were validated by numerical simulations based on exact diagonalization of matrices or Dyson's log-gas method for generating eigenvalues.

Our focus here has been on ensembles of complex random matrices. It is known that in the rotationally invariant cases the ensembles comprising real and quaternion matrices admit joint eigenvalue densities which involve product of a determinant and a Pfaffian~\cite{KP2009,KP2011b}. The antisymmetric kernels involved in the Pfaffian being respectively sgn$(\lambda-\mu)/2$ and $-\delta'(\lambda-\mu)$ in the orthogonal and symplectic ensembles~\cite{KP2009,KP2011b,Kieburg2012b}. Therefore, it is possible to write down the results for gap probabilities and densities of extreme eigenvalues for these ensembles using the type II ensemble results of section~\ref{SecType2}. As a matter of fact, even for orthogonal-unitary and symplectic-unitary crossovers the joint density of eigenvalues has the type II structure~\cite{Mehta2004,KP2009,KP2011b}, therefore results for extreme eigenvalues, in principle, can be worked out. However, in the case of noninvariant ensembles involving real or quaternion matrices, unavailability of group integral formulae similar to those in the unitary group leads to serious complications in obtaining exact results. In these cases one has to deal with hypergeometric functions of matrix arguments, Jack or Zonal polynomials. A more tractable approach is based on supersymmetry, using which explicit answers for extreme eigenvalues have been obtained for the correlated Laguerre-Wishart case~\cite{WG2014,WKG2015,AGKWW,WAGKW}. Moreover, very recently it has been applied to obtain joint density and correlation functions for eigenvalues of the correlated Jacobi ensemble~\cite{WWKK2015}. It will be an interesting but challenging task to explore using supersymmetry exact solutions for the real or quaternion counterparts of the noninvariant ensembles (other than Laguerre-Wishart) considered in this work, and look for the corresponding large-$n$ universal results.

\appendix

\section{Derivation of matrix probability density for correlated Cauchy-Lorentz ensemble}
\label{AppCL}

We derive below the matrix probability density for correlated Cauchy-Lorentz ensemble starting from the matrix model~\eqref{Ratio}. The required density can be obtained as
\begin{equation}
P(\bH)= \int d[\bW_A]\int d[\bW_B] \delta\big(\bH-\bW_A\bW_B^{-1}\big) |\bW_A|^{\alpha}e^{-\tr \bSg_A^{-1}\bW_A}|\bW_B|^{\beta}e^{-\tr \bSg_B^{-1}\bW_B}\Theta(\bW_A)\Theta(\bW_B).
\end{equation}
The delta function with matrix argument in the above equation represents the product of delta functions with scalar arguments, one for each independent real and imaginary component of the matrix argument. Also $\Theta(\bA)$ represents the matrix theta function, and requires the matrix $\bA$ to be positive definite ($\bA>0$) or more generally nonnegative-definite ($\bA\ge0$) for a non-vanishing result.

Employing the transformation $\bW_A\rightarrow \bW_A \bW_B$, we get
\begin{align}
\nonumber
P(\bH)&= \int d[\bW_A]\int d[\bW_B] \delta\big(\bH-\bW_A\big) |\bW_A|^{\alpha}e^{-\tr (\bSg_A^{-1}\bW_A+\bSg_B^{-1})\bW_B}|\bW_B|^{\alpha+\beta+n}\Theta(\bW_A)\Theta(\bW_B)\\
&=|\bH|^{\alpha}\,\Theta(\bH) \int d[\bW_B] \,e^{-\tr (\bSg_A^{-1}\bH+\bSg_B^{-1})\bW_B}|\bW_B|^{\alpha+\beta+n}\Theta(\bW_B)
\end{align}
In the second line above, we performed the trivial delta function integral over $\bA$. Now $\bB$-integral can be easily performed to yield
\begin{align}
\nonumber
P(\bH)&\propto |\bH|^{n_A-n}  |\bSg_A^{-1}\bH+\bSg_B^{-1}|^{-(\alpha+\beta+2n)}\,\Theta(\bH)\\
&\propto |\bH|^{\alpha}|\1+\bSg_B\bSg_A^{-1}\bH|^{-(\alpha+\beta+2n)}\,\Theta(\bH),
\end{align}
which is same as Eq.~\eqref{CL2Corr}.

\section{Derivation of matrix probability density for correlated Jacobi-MANOVA ensemble}
\label{AppMANOVACorr}

The density of $\bH$ as in Eq.~\eqref{MANOVACorr} can be derived using
\begin{align}
\nonumber
P(\bH)\propto \int d[\bW_A]\int d[\bW_B] \delta\big(\bH-\bW_A(\bW_A+\bW_B)^{-1}\big) |\bW_A|^{\alpha}e^{-\tr \bSg_A^{-1}\bW_A}|\bW_B|^{\beta}e^{-\tr \bSg_B^{-1}\bW_B}\\
\times \Theta(\bW_A)\Theta(\bW_B).
\end{align}
Employing the transformation $\bW_B\rightarrow \bW_B\bW_A$ leads to
\begin{align}
\nonumber
P(\bH)\propto &\int d[\bW_A]\int d[\bW_B]\, \delta(\bH-(\1+\bW_B)^{-1}) |\bW_A|^{\alpha+\beta+n}e^{-\tr \bSg_A^{-1}\bW_A}|\bW_B|^{\beta}e^{-\tr \bSg_B^{-1}\bW_B\bW_A}\\
&\times \Theta(\bW_A)\Theta(\bW_B).
\end{align}
The $\bW_A$-integral gives
\begin{equation}
P(\bH)\propto \int d[\bW_B]\, \delta(\bH-(\1+\bW_B)^{-1}) |\bW_B|^{\beta}|\bSg_A^{-1}+\bSg_B^{-1} \bW_B|^{-(\alpha+\beta+2n)}\Theta(\bW_B).
\end{equation}
Taking $(\1+\bW_B)^{-1}=\bY$ yields
\begin{equation}
P(\bH)\propto \int d[\bY]\, \delta(\bH-\bY)  |\bY|^{-2n}|\bY^{-1}-\1|^{\beta}|\bSg_A^{-1}+\bSg_B^{-1} (\bY^{-1}-\1)|^{-(\alpha+\beta+2n)}\Theta(\bY)\Theta(\bY^{-1}-\1),
\end{equation}
and then we have the trivial delta-function integral giving
\begin{equation}
P(\bH)\propto  |\bH|^{-2n}|\bH^{-1}-\1|^{\beta}|\bSg_A^{-1}+\bSg_B^{-1} (\bH^{-1}-\1)|^{-(\alpha+\beta+2n)}\Theta(\bH)\Theta(\bH^{-1}-\1).
\end{equation}
Note that $\bH^{-1}-\1>0$ implies $\bH<\1$. Now, carrying out some readjustments we obtain
\begin{align}
\nonumber
P(\bH)&\propto |\bH|^{\alpha} |\1-\bH|^{\beta}|\bSg_B^{-1}+(\bSg_A^{-1}-\bSg_B^{-1})\bH|^{-(\alpha+\beta+2n)}\Theta(\bH)\Theta(\1-\bH)\\
& \propto |\bH|^{\alpha} |\1-\bH|^{\beta}|\1+(\bSg^{-1}-\1)\bH|^{-(\alpha+\beta+2n)}\Theta(\bH)\Theta(\1-\bH),
\end{align}
where $\bSg^{-1}=\bSg_B \bSg_A^{-1}$.

\section{Derivation of matrix probability density for correlated Bures-Hall ensemble}
\label{AppBH}

We derive here the density for correlated Bures-Hall ensemble. Introducing the Fourier representation for delta function $\delta\big(\bH-\bW(\1+\bSg^{1/2}\bG^2\bSg^{1/2})^{-1}\big)$ using a matrix $\bK$, we have
\begin{equation}
P(\bH)\propto \int d[\bK]\int d[\bG] \int d[\bW]e^{i\tr \bK\bH}e^{-i\tr \bK\bW(\1+\bSg^{1/2}\bG^2\bSg^{1/2})^{-1}}|\1+\bSg \bG^2|^{-(\eta+n)}e^{-\tr\bSg^{-1}\bW}|\bW|^{\eta}\Theta(\bW).
\end{equation}
Integral over $\bW$ leads to
\begin{equation}
P(\bH)\propto \int d[\bK]\int d[\bG] e^{i\tr \bK\bH}|\1+\bSg \bG^2|^{-(\eta+n)}\,|\bSg^{-1}+i  (\1+ \bSg^{1/2}\bG^2\bSg^{1/2})^{-1}\bK|^{-(\eta+n)}.
\end{equation}
Considering $\bG\rightarrow \bSg^{-1/2}\bG\bSg^{1/2}$ and employing some readjustments we obtain
\begin{equation}
P(\bH)\propto \int d[\bK]\int d[\bG] e^{i\tr \bK\bH}|\bSg^{-1} +\bG^2|^{-(\eta+n)}|\1+i(\bSg^{-1}+\bG^2)^{-1} \bK|^{-(\eta+n)}.
\end{equation}
We now use the transformation $\bK\rightarrow (\bSg^{-1}+ \bG^2)\bK$ and get
\begin{equation}
P(\bH)\propto \int d[\bK]\int d[\bG] e^{i\tr \bK\bH (\bSg^{-1}+ \bG^2)}|\bSg^{-1}+ \bG^2|^{-\eta}|\1+i \bK|^{-(\eta+n)}.
\end{equation}
Ingham-Siegal-Fyodorov formula~\cite{Fyodorov2002} can be used to solve the $\bK$-integral, and on simplification we have
\begin{equation}
\label{HGint}
P(\bH)\propto e^{-\tr\bSg^{-1}\bH}|\bH|^{\eta}\,\Theta(\bH)\int d[\bG] e^{-\tr \bG^2\bH}.
\end{equation}
The $\bG$-integral yields result which depends only on the eigenvalues of $\bH$ as~\cite{OSZ2010}
\begin{equation}
\label{Gint}
\int d[\bG] e^{-\tr \bG^2\bH}\propto \frac{1}{\Delta_{+}(\{\lambda\})}\prod_{j=1}^n \lambda_j^{-1/2}.
\end{equation}
This can be proved using the result from correlated Gauss-Wigner case, with the replacements $\bH\rightarrow \bG$ and $\bSg^{-2}\rightarrow \bH$ in Eq.~\eqref{WigCorr}.
We have, with $g_j$ representing the eigenvalues of $\bG$,  
\begin{align}
\nonumber
\int d[\bG] e^{-\tr \bG^2\bH}&\propto \int dg_1\cdots \int dg_n \Delta^2(\{g\})\int\,d\mu(\bU)e^{-\tr \bH \bU^\dag \bG^2\bU}\\
\nonumber
&\propto \int dg_1\cdots \int dg_n \Delta^2(\{g\}) \frac{|e^{-\lambda_j g_k^2}|_{j,k=1,...,n}}{\Delta(\{\lambda\})\Delta(\{g^2\})}\\
&\propto \frac{1}{\Delta(\{\lambda\})}\int dg_1\cdots \int dg_n \frac{\Delta(\{g\})}{\Delta_+(\{g\})} |e^{-\lambda_j g_k^2}|_{j,k=1,...,n}.
\end{align}
We used above HCIZ formula to perform the unitary group integral. Now, the integrals over $g_j$ follow using the normalization factor result in Eq.~\eqref{WigCorrNorm}. We have
\begin{align}
\nonumber
\int d[\bG] e^{-\tr \bG^2\bH}&\propto \frac{1}{\Delta(\{\lambda\})}\prod_{j=1}^n \lambda_j^{-1/2}\cdot\prod_{j>k}\frac{\lambda_j^{-1}-\lambda_k^{-1}}{\lambda_j^{-1}+\lambda_k^{-1}},
\end{align}
which gives on little simplification Eq.~\eqref{Gint}. Therefore, the matrix probability density for $\bH$ may be written as in Eq.~\eqref{BHCorr} with the identification $\alpha=\eta-1/2$.



\end{document}